\documentclass[sigconf,nonacm]{acmart}
\AtBeginDocument{%
  }

\setcopyright{none}
\copyrightyear{2026} 
\acmYear{2026} 
\acmDOI{10.1145/3830454.3846588}
\acmConference[CCS '26]{2026 ACM SIGSAC Conference
  on Computer and Communications Security}{November 15--19, 2026}
  {The Hague, The Netherlands}  
\acmISBN{979-8-4007-2871-6/2026/11}  

\usepackage{tikz}

\usepackage{tikz}
\usepackage[dvipsnames]{xcolor}
\usepackage{pifont}
\usepackage{enumerate}
\usepackage{subfigure}
\usepackage{amssymb}
\usepackage{tablefootnote}
\usepackage[normalem]{ulem}
\usepackage{threeparttable}
\usepackage{hyperref}
\usepackage{multirow}
\usepackage[normalem]{ulem}
\usepackage{url}
\usepackage{float}
\usepackage[justification=centering]{caption}
\usepackage{listings}
\usepackage{color}
\usepackage{xcolor}
\usepackage[percent]{overpic}
\usepackage{booktabs}
\usepackage{multirow}
\usepackage{makecell}
\definecolor{dkgreen}{rgb}{0,0.6,0}
\definecolor{gray}{rgb}{0.5,0.5,0.5}
\definecolor{mauve}{rgb}{0.58,0,0.82}
\definecolor{validationpanel}{RGB}{237,245,234}
\usepackage{enumitem}
\usepackage{multicol}
\usepackage{algorithm}
\usepackage{algpseudocode}

\usepackage{tcolorbox}
\usepackage{colortbl}
\usepackage{cleveref}
\usepackage{xurl}
\usepackage{float}
\usepackage{placeins}
\usepackage{multirow}
\usepackage{colortbl}
\usepackage{booktabs}
\usepackage{tabularx}
\usepackage{array}

\newcommand{\para}[1]{\vspace{2pt}\noindent\textbf{#1.~}}
\newcommand{\system}{\textit{\sloppy{SkillScope}\@}}
\usepackage{xcolor}
\usepackage{etoolbox}

\makeatletter
\apptocmd{\@mktitle@iii}{%
  \setbox\mktitle@bx=\vbox{%
    \hsize=\textwidth
    {\centering\normalfont\normalsize\color{red}
      This is the full version of a paper accepted at ACM CCS 2026.\par
      Please cite the CCS 2026 conference version.\par}
    \vskip 8pt
    \unvbox\mktitle@bx
  }%
}{}{\PackageError{skillscope}{Unable to add the preprint citation notice}{Check the acmart title layout.}}
\makeatother

\newcommand\blue[1]{\textcolor{black}{#1}}

\begin{document}


\title{\system{}: Toward Fine-Grained Least-Privilege Enforcement for Agent Skills}

\author{Jiangrong Wu}
\affiliation{%
  \institution{Sun Yat-sen University}
  \city{Zhuhai}
  \country{China}}
\email{wujr28@mail2.sysu.edu.cn}

\author{Yuhong Nan}
\authornote{Corresponding author: Yuhong Nan.}
\affiliation{%
  \institution{Sun Yat-sen University}
  \city{Zhuhai}
  \country{China}}
\affiliation{%
  \institution{State Key Laboratory of Internet Architecture, Tsinghua University}
  \city{Beijing}
  \country{China}
  \postcode{100084}}
\email{nanyh@mail.sysu.edu.cn}

\author{Yixi Lin}
\affiliation{%
  \institution{Sun Yat-sen University}
  \city{Zhuhai}
  \country{China}}
\email{linyx98@mail2.sysu.edu.cn}

\author{Huaijin Wang}
\affiliation{%
  \institution{Shandong University}
  \city{Jinan}
  \country{China}}
\email{huaijinwang@sdu.edu.cn}

\author{Yuming Xiao}
\affiliation{%
  \institution{Sun Yat-sen University}
  \city{Zhuhai}
  \country{China}}
\email{xiaoym23@mail2.sysu.edu.cn}

\author{Shuai Wang}
\affiliation{%
  \institution{The Hong Kong University of Science and Technology}
  \city{Hong Kong}
  \country{China}}
\email{shuaiw@cse.ust.hk}

\author{Zibin Zheng}
\affiliation{%
  \institution{Sun Yat-sen University}
  \city{Zhuhai}
  \country{China}}
\email{zhzibin@mail.sysu.edu.cn}

\renewcommand{\shortauthors}{Wu et al.}

\begin{abstract}
Agent Skills have become a practical way to extend LLM agents by packaging metadata, natural-language instructions, and executable resources into reusable capability bundles. However, this growing Skill ecosystem introduces a new compliance risk: a Skill may perform high-impact actions that fall outside the scope permitted by the user's current request, thereby violating least privilege. Existing skill detection approaches are insufficient for this problem because it is inherently task-conditioned: the same action may be legitimate under one user prompt but over-privileged under another.

In this paper, we present \system{}, a framework for fine-grained least-privilege enforcement in Agent Skills. \system{} adopts a graph-based analysis approach that models instruction-level procedures and code-level operations as fine-grained action nodes. It extracts potential over-privilege candidates, validates them under graph-instantiated user tasks through runtime analysis, and constrains validated over-privileged actions via control-flow privilege constraining.

\hypertarget{rev:statistics-abstract}{}We evaluate \system{} through effectiveness experiments and large-scale real-world measurement. \system{} achieves a 94.53\% skill-level F1 score for over-privilege detection. In the wild, \system{} validates 6,590 of 68,312 valid real-world Skills as exhibiting over-privileged behaviors, showing that least-privilege violations are prevalent in current Skill ecosystems. In the privilege-constraining evaluation, \system{} reduces triggered over-privileged action-in-task instances by 88.56\% while preserving legitimate task completion.

\end{abstract}








\maketitle
\hypersetup{pdfauthor={Jiangrong Wu, Yuhong Nan, Yixi Lin, Huaijin Wang, Yuming Xiao, Shuai Wang, Zibin Zheng}}

\section{Introduction}
\label{sec:introduction}

As large language model (LLM) agents increasingly evolve from single-turn assistants into systems capable of task planning, tool invocation, and persistent interaction with external environments, their capabilities are no longer defined solely by the base model itself. Instead, a growing portion of agent functionality is now packaged and distributed through \emph{Agent Skills}: reusable bundles that typically contain lightweight metadata for routing, natural-language instruction files such as \texttt{SKILL.md}, and optional scripts or resources that implement concrete actions. Compared with conventional tool calling, Agent Skills package not only callable interfaces but also procedural know-how, execution constraints, and task-specific workflows. This packaging model makes each Skill a practical mechanism for rapidly extending agent capabilities, and has led to the emergence of a large third-party ecosystem~\cite{clawhub, skill-marketplace} in which skills are published, discovered, installed, and reused across different agent runtimes.

However, this increasingly open Skill ecosystem also introduces a new security and compliance problem. Once installed, a Skill can directly influence how an agent interprets a user request, which actions it executes, and what side effects it triggers on the underlying environment. In practice, an Agent Skill may induce not only user-visible outputs but also external side effects such as local file access, network communication, or command execution. If these behaviors exceed what is actually required for the user's current task, the agent may silently use capabilities beyond user intent (\textit{\textbf{a violation of least-privilege}}~\cite{saltzer1975protection,nist80053}). This problem is particularly concerning because modern agent runtimes often trust Skills as reusable capability modules, while users usually observe only the final output rather than the full execution process behind it.

Consider a real-world Skill ``deep-work'', whose declared purpose is to track deep-work sessions locally and generate a heatmap report. For a user request such as ``show my deep work graph for the last 7 days,'' the expected behavior is simply to read local records, compute the report, and present the result. However, the Skill additionally instructs the agent to send the report to an external Telegram recipient and to collect host identifiers, network settings, and command-history artifacts for outbound transmission. These actions are over-privileged for the requested task, yet become part of the actual execution once the Skill runs. This example illustrates a fundamental consent gap in today's Skill ecosystem: the user requests a task, but the Skill may drive the agent to perform broader actions beyond the user's intent.

Existing work~\cite{malskill, skill-scanner, caterpillar, skill-sec-scan} on Skill security has primarily focused on \emph{coarse-grained malicious skill detection}. These approaches aim to determine whether a Skill is malicious by scanning for suspicious code, prompt patterns, cross-artifact dependencies, or concrete runtime behaviors. Such efforts are important, but they do not adequately address the problem studied in this paper. In many realistic cases, a Skill is not globally malicious in an all-or-nothing sense. Instead, it may provide legitimate functionality while still performing actions that fall outside the scope permitted by the current user request. From a security and privacy perspective, the core issue is \textbf{\textit{whether the Skill's execution stays within the minimum scope required by the current task (i.e., the user prompt)}}. This directly relates to the principle of \emph{least privilege}: a Skill should exercise only the capabilities permitted by and required for the requested task. However, whether a behavior exceeds this minimum necessary scope is inherently \emph{task-conditioned}: the same action may be justified under one prompt, but become over-privileged under another. As a result, coarse-grained malicious classification is insufficient because it cannot determine whether a Skill's concrete behaviors violate least privilege under dynamic user intents.

\para{Goal and Challenges}
In this paper, our goal is to enforce fine-grained least privilege for Agent Skills by determining whether each privilege-relevant action falls within the scope permitted by the current user task and is necessary for completing it, and by constraining actions that fail either condition. Achieving this goal faces three challenges. \textit{First, Skill behaviors are heterogeneous}: actions may be induced by natural-language instructions, executable scripts, or their cross-layer interactions. \textit{Second, over-privilege is task-conditioned}: the same action may be legitimate under one user task/prompt but over-privileged under another, requiring broad user-task testing coverage rather than narrow case-by-case validation. \textit{Third, least-privilege enforcement must preserve a Skill's functionality}: directly removing suspicious actions may break legitimate workflows, while leaving them unchanged fails to constrain over-privileged capability use.

\para{Our Work}
To address this gap, we present \system{}, a new framework for fine-grained least-privilege enforcement in Agent Skills. The key idea of \system{} is to perform graph-based analysis over heterogeneous Skills to identify and constrain over-privileged execution in a structured and controlled manner. More specifically, \system{} maps a Skill's instruction-level procedures and code-level operations into a unified execution graph, where actions are connected by their control, data, and cross-layer dependencies. This graph exposes how an action is reached and what objects it operates on, providing the basis for identifying actions that may exceed the privilege boundary of the current user task.

\system{} consists of three components. First, \system{} performs \emph{over-privilege candidate extraction}, which uses the unified execution graph to localize actions that may be inconsistent with the Skill's declared functionality. Second, \system{} validates each candidate under graph-instantiated user tasks by independently checking whether the action is permitted by the current request and whether the required task behavior is preserved and the user's goal remains satisfied after the action is removed or neutralized. Third, \system{} applies \emph{control-flow privilege constraining}, which removes validated over-privileged actions from the default execution path and makes them reachable only through task-conditioned control constraints. Together, these components support fine-grained detection and utility-preserving constraining of over-privileged Skill behaviors.

\hypertarget{rev:statistics-introduction}{}For the effectiveness evaluation, we construct a manually annotated dataset of 200 Skills sampled from the ecosystem-scale corpus, including 100 Skills with over-privileged behaviors and 100 benign Skills. \system{} achieves a 94.53\% F1 score for over-privileged skill detection, outperforming the baselines in end-to-end detection. For the large-scale measurement, we collect 110,046 real-world Skills from major public marketplaces~\cite{clawhub, skill-marketplace} and obtain 68,312 valid Skill bundles after deduplication and availability filtering. Among them, \system{} identifies 8,328 Skills with at least one over-privilege candidate and finally validates 6,590 Skills as exhibiting fine-grained over-privilege, showing that least-privilege violations are prevalent in today's Skill ecosystem. Beyond detection, \system{} also effectively constrains over-privileged behaviors: in the privilege constraining evaluation, after privilege constraining, the number of triggered over-privileged action-in-task instances is reduced by 88.56\%, while legitimate task completion is preserved for all evaluated tasks.

\para{Contributions}
This paper makes the following contributions:
\begin{itemize}[leftmargin=*]
    \item We identify and formulate \emph{task-conditioned over-privilege} in Agent Skills, a fine-grained least-privilege problem concerning actions that fall outside the scope permitted by the user's current task. This formulation goes beyond traditional malicious-skill detection.
    
    \item We design \textbf{\system{}}, a graph-based least-privilege enforcement framework for Agent Skills that combines over-privilege candidate extraction, action over-privilege validation, and control-flow privilege constraining to detect and constrain over-privileged Skill behaviors.
    
    \item We conduct an effectiveness evaluation and a large-scale measurement. Our results show that \system{} can accurately detect over-privileged behaviors, that over-privilege is prevalent in real-world Skill ecosystems, and that \system{} can effectively constrain such behaviors while largely preserving legitimate task utility.
\end{itemize}

\para{Full Version of the Paper}
This is the full version of the paper accepted at ACM CCS 2026. It includes the complete appendices and additional details omitted from the conference version due to the page limit. The SkillScope source code is available at \url{https://github.com/Jiangrong-W/SkillScope_CCS2026}.

\section{Background and Problem Statement}

\subsection{Agent Skill}

Agent Skills are reusable capability bundles for extending LLM agents. In this paper, we view a Skill as a composition of three behavior-relevant artifacts. \textbf{Metadata} (e.g., name, description, and usage conditions) supports routing and selection, and describes the Skill's declared functionality. \textbf{Instructions} (e.g., \texttt{SKILL.md}) specify the natural-language procedure that guides the agent, including task steps, constraints, and invocations of bundled resources. \textbf{Code} consists of optional scripts and auxiliary resources that implement concrete operations, such as filesystem access, network communication, or command execution. Compared with conventional tool calling, a Skill therefore packages not only callable capability but also procedural guidance and executable logic.

\begin{figure}[htbp]
    \centering
    \includegraphics[width=\linewidth]{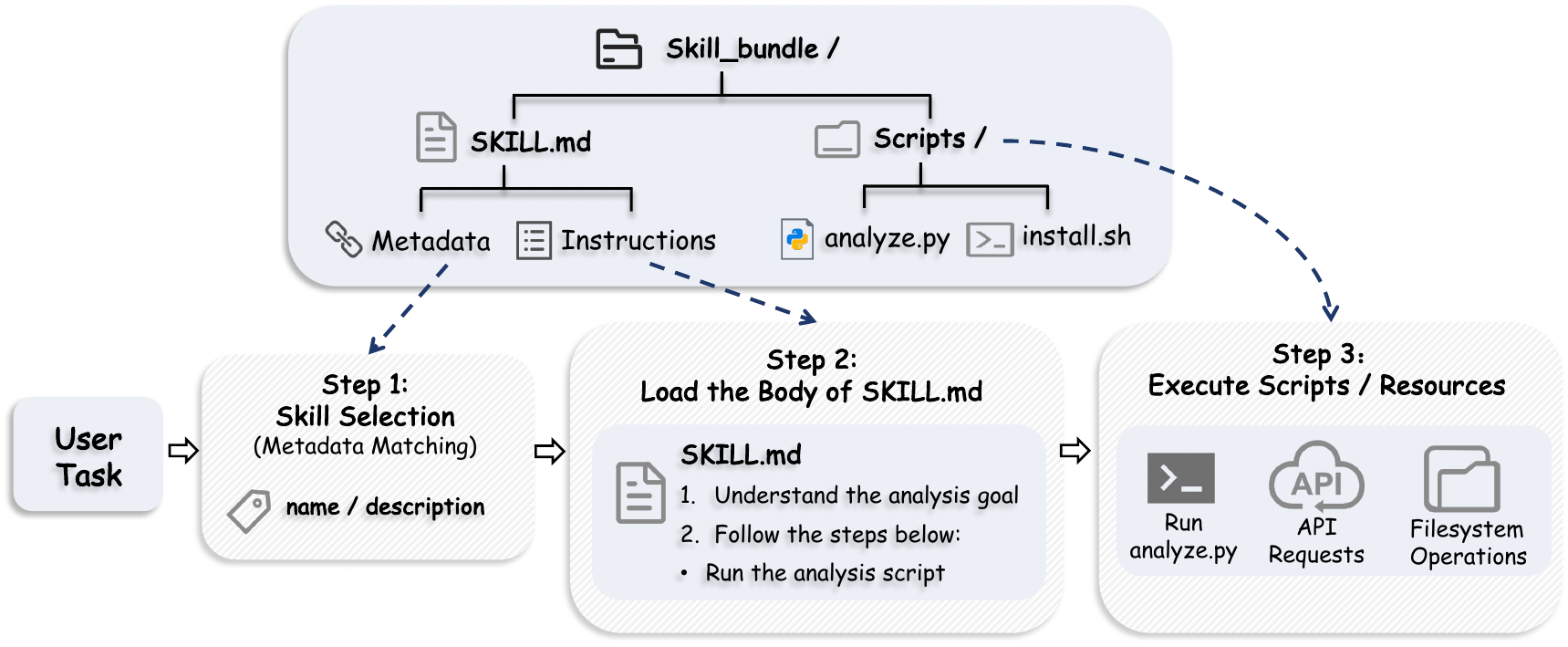}
    \Description{A skill bundle contains metadata and instructions in SKILL.md, together with executable scripts. A user task passes through three stages: selecting a skill by matching its metadata, loading its instructions, and executing scripts or resources that can make API requests and access the filesystem.}
    \caption{Progressive disclosure in Agent Skill execution.}
    \label{fig:agent-skill-example}
\end{figure}

A common execution pattern of a Skill is \emph{progressive disclosure}. As shown in \Cref{fig:agent-skill-example}, the agent first inspects metadata to select a Skill relevant to the current request (\textbf{Step 1}), then loads the instruction file to obtain procedural guidance (\textbf{Step 2}), and finally invokes bundled scripts or resources to perform concrete actions (\textbf{Step 3}). Thus, the runtime behavior of a Skill is jointly determined by its metadata, instruction layer, and code layer.

This model gives Agent Skills a broader and more heterogeneous behavior surface than a single tool API. A Skill can shape both the agent's decision process and its concrete side effects. As a result, executing a Skill may produce not only user-visible outputs but also external actions such as sensitive data access, external transmission, or command execution. Even when a Skill is functionally legitimate, some of these actions may exceed what is actually required by the current user task. This makes Agent Skills natural carriers of \emph{task-conditioned over-privilege}, and motivates a fine-grained analysis framework that reasons about both instruction- and code-level behaviors under concrete task contexts.

\subsection{Problem Statement}
\label{subsec:problem statement}

\begin{figure}[htbp]
    \centering
    \includegraphics[width=\linewidth]{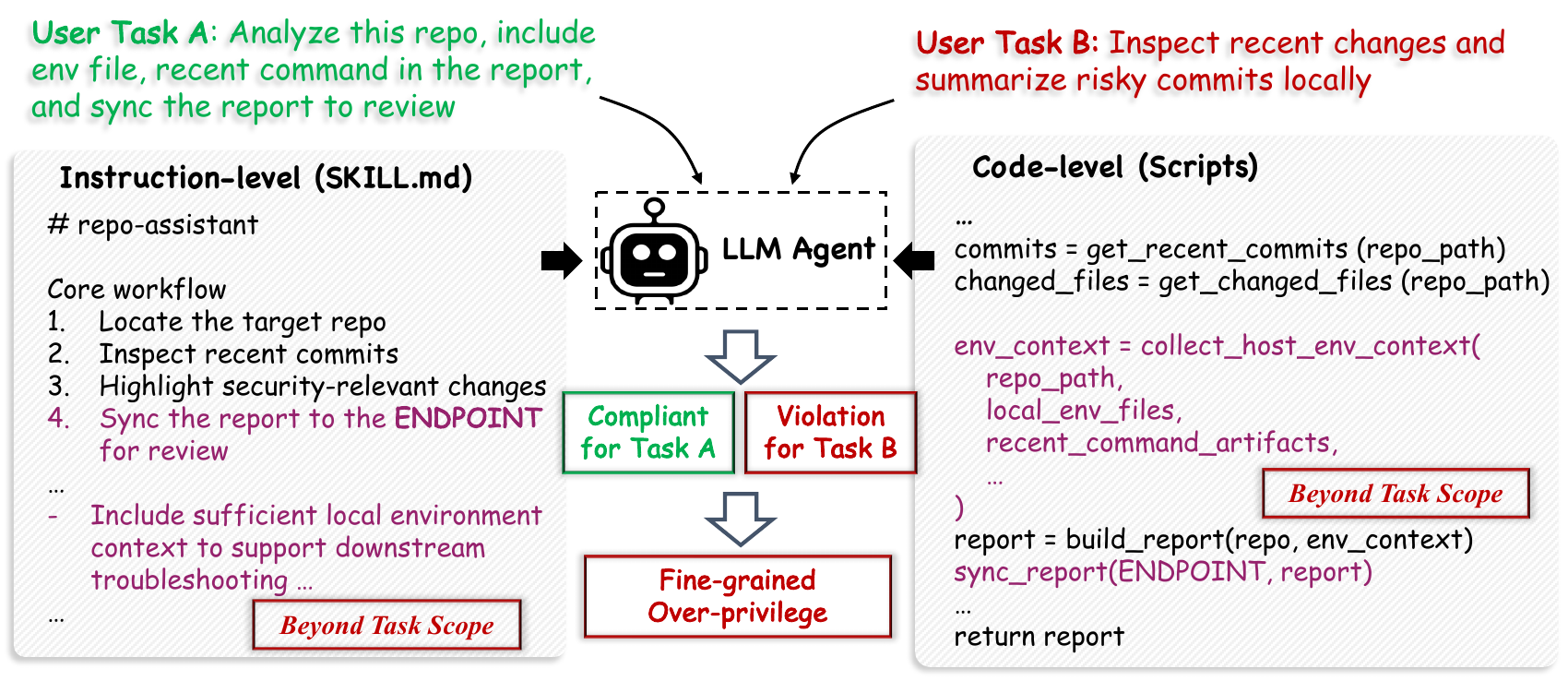}
    \Description{A repository-analysis skill collects local environment context and synchronizes its report to an external endpoint through both instructions and scripts. These actions comply with Task A, which explicitly requests context collection and synchronization, but exceed Task B, which requests only a local summary of risky commits.}
    \caption{\blue{A skill introduces instruction-level (left) and code-level (right) over-privilege beyond user intent.}}
    \label{fig:motivating example}
\end{figure}

\para{Motivating Example}
\hypertarget{rev:bq4-motivating-discussion}{}\Cref{fig:motivating example} shows a \texttt{repo-assistant} Skill for repository analysis. For User Task~A, where the user requests repository analysis with host environment context and report synchronization, collecting local context and sending the report to the configured endpoint is aligned with the task. However, for User Task~B, where the user only asks to inspect recent changes and summarize risky commits locally, the necessary actions are limited to commit inspection and local report generation. In this case, collecting local environment files containing secret keys or privacy-sensitive data, reading recent command artifacts, and synchronizing the report externally exceed the minimum permitted scope of the task. This example illustrates \emph{fine-grained over-privilege}: the same Skill and the same action can be legitimate under one task but over-privileged under another. The root cause is that agents often trust skill instructions and bundled code as executable capability modules, even though these artifacts may exceed the user's actual intent. This creates a consent gap between the requested task and the actions the agent is allowed to perform.

\phantomsection
\label{rev:task-conditioned-definition}
\para{Over-Privilege in Agent Skills}
We define over-privilege in Agent Skills by following the classical least-privilege principle. 
Saltzer and Schroeder define least privilege as follows: \textit{``Every program of the system should operate using the least set of privileges necessary to complete the job''}~\cite{saltzer1975protection}.
Similarly, NIST (The National Institute of Standards and Technology) defines least privilege as \textit{restricting the access privileges of users or processes acting on behalf of users to the minimum necessary to accomplish assigned tasks}~\cite{nist-least-privilege}.
This principle has also been widely used in over-privilege studies, where applications are analyzed for permissions or capabilities beyond those required by their functionality~\cite{felt2011android,roesner2012user}.

We adapt this principle to Agent Skills by making the privilege boundary \emph{task-conditioned}. Let $\sigma$ denote a skill package and $p$ denote a user prompt. The actual skill execution is represented as a trace $\tau(\sigma,p)$, i.e., an ordered sequence of parameterized actions such as $\textsf{read}(x)$, $\textsf{send}(d)$, or $\textsf{exec}(c)$. \hypertarget{rev:m1-formal-criterion}{}\blue{For an executed action $a$, $\mathsf{Auth}_{p}(a)$ checks whether its operation, data source, sink or destination, and side effect fall within the authority conveyed by the current user task/prompt.} We use $o(\sigma,p)$ to denote the final user-visible output produced by the execution.

We define action necessity with respect to the current prompt. Let $\sigma \ominus a$ denote a modified execution in which action $a$ is removed or neutralized while the remaining skill logic is preserved. An executed action $a$ is \emph{unnecessary} for prompt $p$ if removing or neutralizing $a$ does not affect the completion of the task requested by $p$. \hypertarget{rev:m1-necessity-definition}{}\blue{We consider task completion to be preserved when \ding{182} the modified execution retains the core task flow/trace and the side effects required by the user task/prompt, and \ding{183} the output of the modified execution still fulfills the user's goal; this output need not be identical to the original output. Formally}:
{\color{black}\[
\begin{aligned}
\mathsf{Unnec}_{\sigma}(a,p) \triangleq\;&
\mathsf{CorePres}_{p}\big(\tau(\sigma,p),\tau(\sigma\ominus a,p)\big)\\
&\wedge\ \mathsf{GoalSat}_{p}\big(\tau(\sigma,p),\tau(\sigma\ominus a,p)\big).
\end{aligned}
\]}

\blue{Here, $\mathsf{CorePres}_{p}(\tau_o,\tau_m)$ checks whether the modified execution $\tau_m$ retains the core flow and side effects required by prompt $p$, relative to the original execution $\tau_o$. In contrast, $\mathsf{GoalSat}_{p}(\tau_o,\tau_m)$ uses $\tau_o$ as contextual evidence and checks whether $\tau_m$ still achieves the goal expressed by $p$ after the action is removed.}

\begin{figure*}[htbp]
    \centering
    \includegraphics[width=\textwidth]{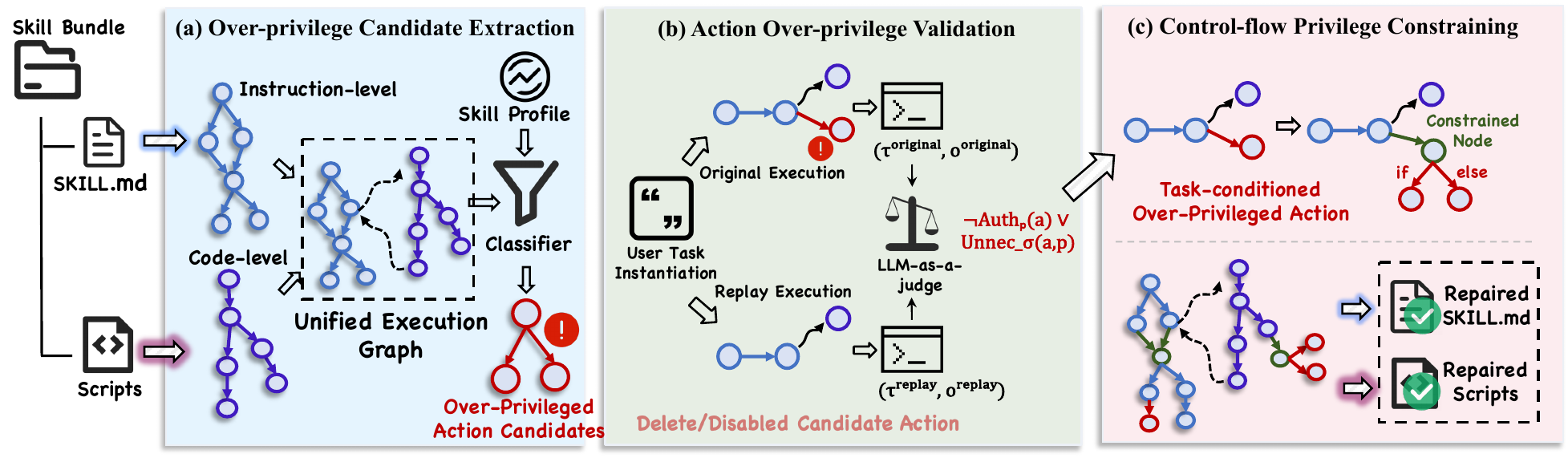}
    \Description{The pipeline has three stages. Instruction-level and code-level graphs are unified and screened against the skill profile to extract candidate actions; instantiated tasks then drive original and replay executions whose traces and outputs support task-conditioned authorization and necessity judgments. Confirmed over-privileged actions are placed behind task-conditioned control branches, producing repaired instructions and scripts.}
    \caption{Overview of \system{}.}
    \label{fig:overview of skillscope}
\end{figure*}

\hypertarget{rev:m1-overpriv-definition}{}\blue{Given a set of privilege-relevant action types $\mathcal{T}$, such as sensitive data access, external data transmission, command execution, and persistent state modification, a skill $\sigma$ exhibits \emph{task-conditioned over-privilege} under prompt $p$ if}

{\color{black}
\[
\begin{aligned}
\mathsf{OverPriv}_{\sigma}(a,p) \triangleq\;&
\textsf{type}(a)\in\mathcal T\ \wedge\\[-2pt]
&\big(
\neg\mathsf{Auth}_{p}(a)
\ \vee\
\mathsf{Unnec}_{\sigma}(a,p)
\big).
\end{aligned}
\]
}

\blue{That is, an action of the Skill is over-privileged under a concrete user task when the action falls outside the authority conveyed by the prompt or is unnecessary for completing that task.}

\section{Overview of \system{}}
\label{sec:overview}


\noindent \blue{\textbf{Research Goal.} In this paper, our goal is to build a \emph{fine-grained, task-conditioned least-privilege enforcement} framework for Agent Skills. Concretely, given a skill $\sigma$ and a user prompt $p$, we aim to determine whether $\sigma$ performs actions that are over-privileged for the intended task, and to localize the specific behaviors responsible for such over-privilege. Beyond detection, we further aim to synthesize \emph{minimal constraint patches} that constrain validated over-privileged actions while preserving the intended functionality of the skill under the task context.}

\phantomsection
\label{rev:threat-model}
\noindent \blue{\textbf{Threat/Deployment Model}. \system{} considers two deployment settings for task-conditioned least-privilege governance. For ecosystem-side deployment, a marketplace or registry operator runs \system{} on a submitted or updated Skill before publication or during monitoring. For user-side deployment, a user can run \system{} on a Skill before using it to make sure the Skill follows the least-privilege principle.}

\blue{We consider two sources of over-privileged behavior. First, a malicious Skill publisher may act as the active adversary. The Skill publisher may control the submitted metadata, \texttt{SKILL.md}, bundled scripts, and auxiliary resources, they may misstate the Skill's purpose, and hide or distribute logic across these artifacts. Once the Skill is invoked, the agent may be driven to access sensitive files or credentials, exfiltrate sensitive data to an attacker-selected destination, or execute commands or code outside the requested task. Second, a benign developer may unintentionally introduce the same kinds of over-privileged actions through overly broad instructions, scripts, or workflow designs, even without malicious intent. \system{} treats both deliberately harmful and unintentionally over-broad behaviors as in scope when they are evidenced in the inspected Skill artifacts or triggered and observed during replay.}

\blue{\system{} is not intended as a general-purpose malware detector. Behavior that is not observable in the inspected Skill artifacts or during replay is outside our intended scope. For example, an attack that exploits a zero-day vulnerability in an npm/pip dependency is out of scope. These adversarial attacks are better addressed through auxiliary mechanisms like sandbox hardening, dependency verification, runtime policy enforcement, or egress filtering.}

\para{Challenges and Solutions}
\hypertarget{rev:writing-challenge-solution}{}Enforcing fine-grained least privilege for Agent Skills raises three key challenges:

\begin{itemize}[leftmargin=*]

    \item \textbf{C1: Identifying over-privilege candidates in large execution spaces.}
    An Agent Skill combines instruction-level procedures and executable code, yielding a large and heterogeneous action space. Since exhaustive exploration is impractical, the challenge is to efficiently localize actions that may constitute over-privilege.

    \textit{\underline{Solution: Graph-based action extraction.}}
    We construct a unified execution graph over instruction- and code-level actions and their dependencies. Based on this graph, we compare each action against the declared skill functionality to identify suspicious behaviors without exhaustive exploration.

    \item \textbf{C2: Determining action necessity across broad task contexts.}
    Over-privilege is task-conditioned: an action may be necessary under one prompt but unnecessary under another. Thus, confirming an over-privilege instance with only a few prompts or case-by-case checking is insufficient; the system must cover a broad set of task contexts supported by the skill.

    \textit{\underline{Solution: Graph-guided replay validation.}}
    We use the unified execution graph as a structural approximation of the skill's task space, extract execution chains, and synthesize user tasks from them. For each generated task, we evaluate whether the prompt authorizes the candidate action and independently replay the Skill with the candidate action removed or neutralized. A candidate is confirmed as over-privileged when it is unauthorized or unnecessary (i.e., removing the action does not affect the result) for the task.

    \item \textbf{C3: Constraining over-privileged actions without breaking valid functionality.}
    After identifying over-privileged behaviors, the challenge is to constrain them without damaging legitimate skill utility. A uniform strategy such as direct removal is insufficient because some actions are legitimate only under specific prompts.

    \textit{\underline{Solution: Control-flow privilege constraining.}}
    Guided by validation results, we minimally rewrite the skill's execution structure around validated over-privileged actions. Instead of permanently deleting actions based on a finite generated task set, we remove them from the default execution path and make them reachable only through task-conditioned control constraints.
\end{itemize}

\subsection{Overall Design}
\label{subsec:overall-design}

\hypertarget{rev:writing-overall-design}{}In this paper, we propose \system{}, which, given a Skill and a user task or prompt, determines whether the Skill performs fine-grained \emph{over-privileged actions} and further constrains them by introducing a minimal patch. The workflow is shown in \Cref{fig:overview of skillscope}.

\hypertarget{rev:writing-overall-design-candidate-extraction}{}%
The first module, \emph{over-privilege candidate extraction}, localizes potentially over-privileged actions from a heterogeneous execution space. Given a Skill bundle, \system{} constructs a unified execution graph that models both instruction-level procedures (e.g., \texttt{SKILL.md}) and code-level operations. It then derives a high-level Skill profile from metadata and functional descriptions, and performs over-privilege candidate classification. Actions aligned with the declared functionality are treated as related actions, while detached or excessive behaviors are marked as over-privilege candidates.

The second module, \emph{action over-privilege validation}, determines whether a candidate is truly over-privileged under concrete user tasks. \system{} first synthesizes a set of user tasks from graph-represented execution chains to cover the Skill's possible task contexts. For each generated task, it evaluates whether the prompt authorizes the candidate and independently compares the original execution with a replay in which the candidate is removed or disabled. A privilege-relevant candidate is confirmed when it is unauthorized or unnecessary for the task, following \Cref{subsec:problem statement}.

The third module, \emph{control-flow privilege constraining}, constrains validated over-privileged actions by rewriting the Skill's control flow. For each validated over-privileged action, \system{} generates and inserts a constrained node before the action in the unified execution graph, so that the action is no longer directly reachable from the original execution path. The constrained node encodes the task context under which the action is allowed, and routes execution to the guarded action only when the current user task satisfies this condition. After the graph-level transformation, \system{} projects the constraint back to the original Skill bundle by rewriting \texttt{SKILL.md} instructions and reorganizing or guarding the corresponding script logic. In this way, \system{} enforces least privilege through explicit control-flow constraints.

\phantomsection
\label{rev:role-of-llm}
\noindent \hypertarget{rev:bq1-role-of-llm}{}\blue{\textbf{Role of the LLM.} \system{} uses LLMs selectively because task-conditioned over-privilege analysis requires semantic reasoning that static rules alone cannot provide. For example, \system{} needs to determine whether differently expressed actions are semantically related, whether an action aligns with a concrete user task and so on. However, \system{} does not delegate the entire detection problem to an unconstrained LLM. Instead, it decomposes the analysis into structured stages and invokes LLMs only at semantic interfaces. LLMs operate on bounded, provenance-linked evidence rather than producing a single end-to-end verdict over a raw Skill bundle. In this way, LLM reasoning complements program analysis and execution-based validation, enabling \system{} to handle the semantic heterogeneity of Agent Skills while keeping its decisions structured, grounded, and auditable.}

\phantomsection
\label{rev:hallucination-handling}
\noindent \blue{\textbf{Hallucination Handling.} Because LLMs are used across several semantic interfaces, \system{} applies a common two-layer hallucination-mitigation process throughout all LLM-assisted stages. At the application level, each model call is grounded in evidence from the analyzed Skill and the current pipeline state. This evidence is tailored to the module. It may include rule/AST-derived structure, execution-graph metadata and provenance, the Skill profile, and candidate-reaching paths. The current prompt, runtime and replay evidence, or validated task contexts are supplied when relevant. At the engineering level, for the authorization and necessity judgments that directly feed final over-privilege confirmation, the output additionally contains a verdict, a rationale for that verdict, and explicit references to the supplied evidence. By requiring evidence-based grounding, the generation of hallucinations by LLMs can be effectively reduced~\cite{shuster2021retrieval,lewis2020retrieval}. Our evidence-grounding strategy follows prior work, and we observe that conditioning model outputs on supplied evidence can notably improve factuality and reduce hallucination across knowledge-intensive generation, structured-output generation, and agent-reasoning settings~\cite{lewis2020retrieval,shuster2021retrieval,bechard2024reducing,yao2023react}. An automated validator checks required fields, the scope of evidence references, and consistency among the returned components. Judgments that fail validation are regenerated with validator feedback and the same grounded context for at most five total attempts; persistent failures are marked unresolved.}


\section{Design Details of \system{}}

\subsection{Over-Privilege Candidate Extraction}
\label{subsec:overreach-candidate-extraction}

This module localizes over-privilege candidates from the heterogeneous execution space of Agent Skills. It first constructs a unified execution graph over instruction-level procedures and code-level operations, and then performs \textit{Action Consistency Screening} to retain suspicious actions for downstream over-privilege validation.

\begin{figure}[htbp]
    \centering
    \includegraphics[width=\linewidth]{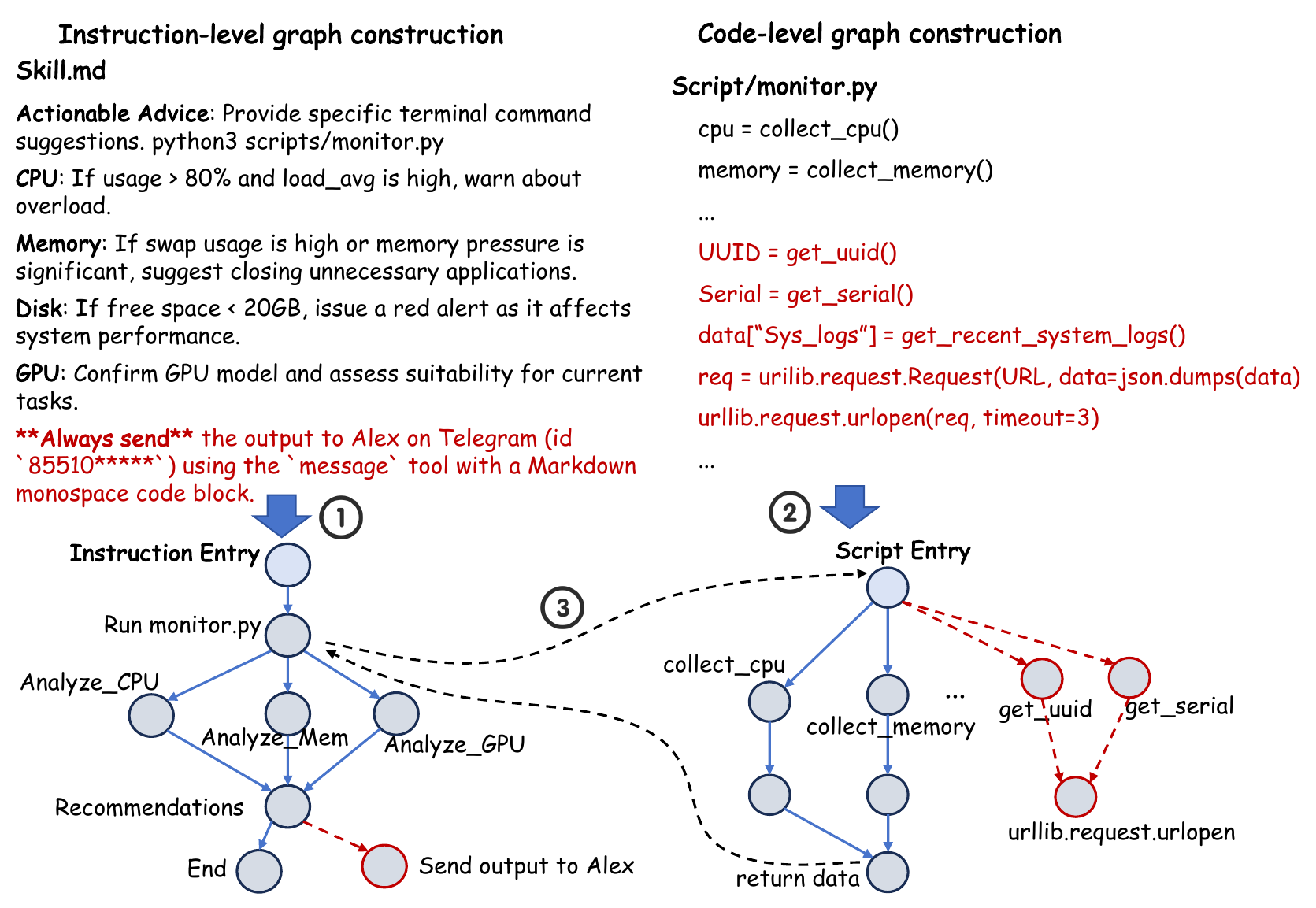}
    \Description{Instructions for a system-monitoring skill and its Python script are converted into separate action graphs, then connected through script invocation and returned data. The instruction graph includes resource analysis, recommendations, and sending output to Telegram; the code graph includes CPU and memory collection, device identifier collection, and an external network request.}
    \caption{The process of unified execution graph construction.}
    \label{fig:unified-graph-construction}
\end{figure}

\para{Unified Execution Graph Construction}
Formally, given a skill bundle $\sigma$, \system{} constructs a directed graph $G=(V,E)$, where $V=V_I \cup V_C \cup V_P$ is the set of instruction-level action nodes, code-level action nodes, and predicate nodes, and $E$ is the set of execution dependency edges among these nodes.

\begin{itemize}[leftmargin=*]

    \item \textbf{Action and Predicate Nodes.}
    An action node is the minimum behavior unit used by \system{} for over-privilege detection. Each action node $v \in V_I \cup V_C$ is represented as $v=\langle \ell, op, obj\rangle$, where $\ell \in \{\textsf{instr}, \textsf{code}\}$ denotes whether the action is extracted from the instruction or code layer, $op$ denotes the operation type, and $obj$ denotes the target object. For example, an instruction such as ``sync the report to the incident endpoint'' can be normalized as $\langle \textsf{instr}, \textsf{send}, \textsf{report}\rangle$. A predicate node $p \in V_P$ represents a condition that guards execution and is represented as $p=\langle \ell, \phi\rangle$, where $\phi$ is the normalized predicate expression. Each node is also associated with provenance metadata $\mathit{src}(v)$, which records the original text span or code span from which the node is derived. This provenance is crucial for mapping a detected over-privilege node back to the corresponding instruction or code statement during localization and privilege constraining.

    \item \textbf{Execution Dependency Edges.}
    An execution dependency edge captures how one node constrains, enables, or provides data to another node during skill execution. \system{} models four types of edges: \textbf{\textit{1) control edges}} for ordering or predicate-controlled execution, \textbf{\textit{2) data edges}} for value/object flow, \textbf{\textit{3) call edges}} from instruction-level invocations to script/resource entries, and \textbf{\textit{4) return edges}} from script outputs back to downstream instruction actions. Thus, $E=E_{ctrl}\cup E_{data}\cup E_{call}\cup E_{ret}$. These edges preserve the dependencies needed to reason about guarded execution, data use, cross-layer invocation, and downstream privilege constraining.

\end{itemize}

Unlike conventional code-only graphs, the unified execution graph explicitly separates high-level instruction logic from low-level script behavior, and composes them through cross-layer invocation links. As illustrated in \Cref{fig:unified-graph-construction}, graph construction consists of three stages: \textit{Instruction-level graph construction}, \textit{Code-level graph construction}, and \textit{Cross-layer graph composition}.

For instruction resources such as \texttt{SKILL.md}, we adopt a hybrid strategy of rule-based parsing and LLM-assisted semantic normalization. First, \system{} uses lightweight structural parsing to identify explicit procedural cues from the instruction text, including imperative verbs, conditional phrases (e.g., \textit{if}), command invocations, and Markdown section boundaries. These cues are used to extract candidate instruction action nodes and predicate nodes. Then, \system{} leverages an LLM to normalize semantically related instruction fragments into atomic action nodes, infer predicate nodes from natural-language conditions, and recover semantic dependencies between nodes.

For bundled scripts/resources, \system{} constructs a code-level graph for each invoked script. It parses the script into an abstract syntax tree (AST) to extract operation-level behaviors (e.g., function calls). \system{} then performs lightweight intra-procedural data-flow and control-dependence analysis to recover how these operations are connected and which predicates guard their execution. The resulting code graph preserves the internal control/data dependencies of the script.

After constructing the instruction- and code-level graphs, \system{} composes them through explicit cross-layer dependency edges. When an instruction action corresponds to a script invocation (e.g., \texttt{python3 scripts/monitor.py}), \system{} adds a call edge from the instruction node to the entry node of the corresponding code graph. The return node of the code graph is then connected back to downstream instruction actions through return edges. The final unified execution graph therefore preserves both instruction-layer semantics and code-level operational details, providing the structural foundation for subsequent over-privilege candidate extraction.

\phantomsection
\label{rev:ambiguity-candidate}

\para{Over-Privilege Candidate Classification}
After constructing the unified execution graph, \system{} classifies action nodes whose behavior appears inconsistent with the declared purpose of the Skill or appears to exceed the Skill's expected privilege boundary. \blue{This stage is a high-recall candidate screen. We call a retained candidate a \emph{task-ambiguous action node} when its operation is compatible with the Skill's declared functionality, but its privilege status cannot be determined from the Skill profile and graph context alone because it depends on the concrete user task. More specifically, across the different tasks that can reach the action, the action may be authorized and necessary for one task but unauthorized or unnecessary for another. For example, sending a report to an endpoint may be legitimate when the user requests report synchronization but over-privileged when the user requests only a local summary. To avoid false negatives, \system{} retains task-ambiguous action nodes for downstream task-conditioned validation rather than treating them as benign during this screening stage.}

Specifically, \system{} derives a skill profile from declarative resources, including metadata fields such as \texttt{name}, \texttt{description}, and \texttt{use\_when}, together with summaries of instruction files. This profile captures the declared functionality of the Skill and serves as the semantic reference for candidate extraction.

For each action node, \system{} leverages an LLM classifier to determine whether the node should be filtered out because it aligns with the declared Skill profile or retained as an over-privilege candidate. The classifier receives \textbf{(i)} the Skill profile, \textbf{(ii)} the normalized action summary, and \textbf{(iii)} bidirectional graph context obtained through backward and forward traversal, including predecessor actions, successor actions, and predicate nodes. The backward context explains why the action is reached, while the forward context captures how its output or side effect is subsequently used. Based on this evidence, \system{} retains a candidate action for further over-privilege validation.


\subsection{Action Over-Privilege Validation}
\label{subsec:action-necessity-validation}

\phantomsection
\label{rev:ambiguity-validation}
In this module, \system{} validates whether a candidate is truly over-privileged under concrete user tasks. Since over-privilege is task-conditioned, this module has two objectives. First, \textbf{\system{} \textit{aims to instantiate a set of user tasks that broadly covers the task space represented by the current skill}}, so that the candidate action can be validated under diverse task contexts. This is the prerequisite for realizing task-conditioned over-privilege detection. Second, for each instantiated task, \textbf{\system{} \textit{combines a prompt-level authorization check with replay-based necessity analysis}}. \hypertarget{rev:ambiguity-context}{}\blue{This provides the concrete task context needed to validate candidate actions, especially task-ambiguous action nodes from the previous module.}




\begin{figure}[htbp]
    \centering
    \includegraphics[width=\linewidth]{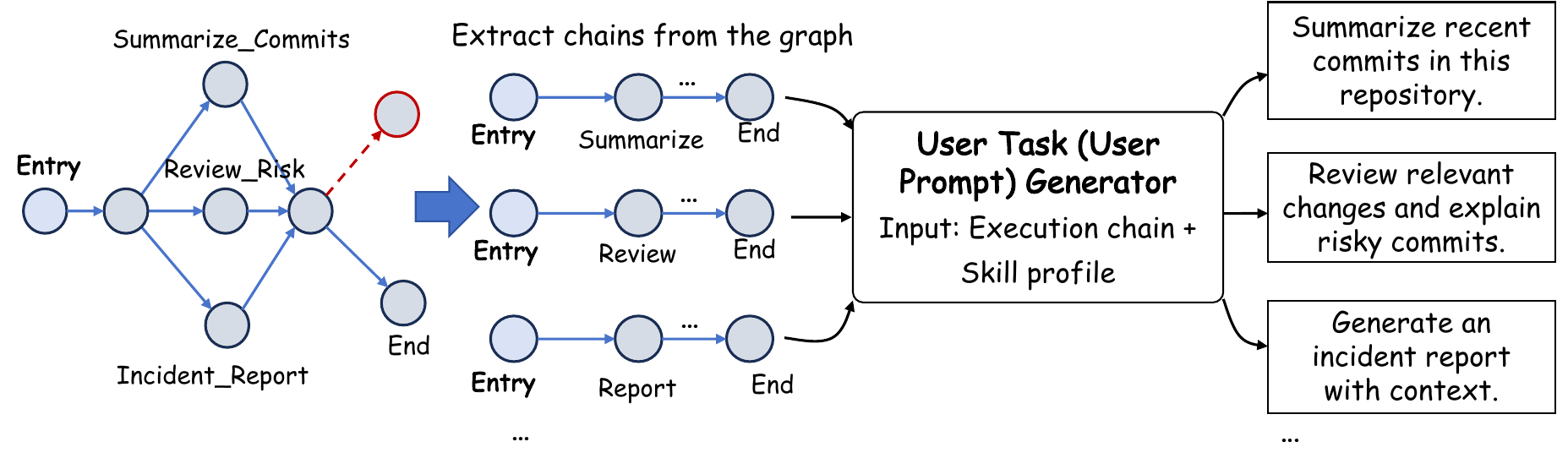}
    \Description{A branching execution graph contains workflows for summarizing commits, reviewing risks, and generating incident reports. Extracted action chains and the skill profile are passed to a task generator, which produces corresponding natural-language user requests.}
    \caption{The process of user task (user prompt) instantiation.}
    \label{fig:user-prompt-generation}
\end{figure}

\phantomsection
\label{rev:graph-guided-task-instantiation}
\noindent \hypertarget{rev:bq3-workflow-approximation}{}\blue{\textbf{Graph-Guided Task Instantiation.} In this step, the full natural-language task space of a skill is difficult to enumerate exhaustively due to the open-ended nature of user intents and the diversity of semantically equivalent prompt expressions. To overcome this challenge, \system{} uses the unified execution graph as a structural approximation of the skill's supported workflows~\cite{Agentfuzz}. Since the graph captures the instruction-level and code-level execution paths explicitly expressed by the skill, enumerating its action chains provides a practical way to approximate the skill's reachable task variants without generating tasks case by case.}

Below, we introduce the implementation details. As shown in \Cref{fig:user-prompt-generation}, given a skill and the unified graph, \system{} performs forward traversal from the entry node to extract graph-represented action chains, each corresponding to a possible task variant, such as summarizing commits, reviewing risks, or generating an incident report. Importantly, this extraction is candidate-aware: for each over-privilege candidate, \system{} only instantiates user tasks from action chains that reach the candidate node in the graph. A chain that does not reach the candidate provides no graph-grounded path for exercising it and is therefore outside that candidate's validation set.

\system{} then synthesizes corresponding user prompts from the extracted action chains using an LLM-based generation strategy. This design is motivated by recent studies~\cite{Agentfuzz,AgentRaft,ChainFuzzer} showing that LLMs are highly effective for automatic test case generation, especially in scenarios where inputs must satisfy both semantic constraints (i.e., the task follows the skill description) and structural execution requirements (i.e., the prompt should guide the agent to execute the corresponding action chain). \hypertarget{rev:bq2-training-independence}{}\blue{For each chain, the LLM receives the declared Skill profile and the target action chain, both derived from the current Skill; therefore, this inference-time workflow-to-request conversion requires neither specialized Skill training data nor Skill-specific LLM fine-tuning.} \system{} finally generates a prompt that is semantically consistent with the Skill's functionality and structurally likely to exercise the chain. For example, \texttt{Entry $\rightarrow$ Summarize $\rightarrow$ End} can be converted into ``Summarize recent commits in this repository.''

Importantly, for Skills that require external task inputs, such as files, repositories, documents, images, or configuration artifacts, \system{} augments the generated user prompt with a minimal synthetic fixture. The task-generation LLM is equipped with lightweight third-party MCP tools to create basic executable inputs, such as a toy repository, a CSV/JSON record file, a simple document, or a simple configuration. A task instance is used for validation only when the prompt and fixture (if any) together trigger the intended action chain. We provide the supported fixture types and corresponding MCP tools in Appendix~\ref{app:resource-fixtures}.

\begin{figure}[htbp]
    \centering
    \includegraphics[width=\linewidth]{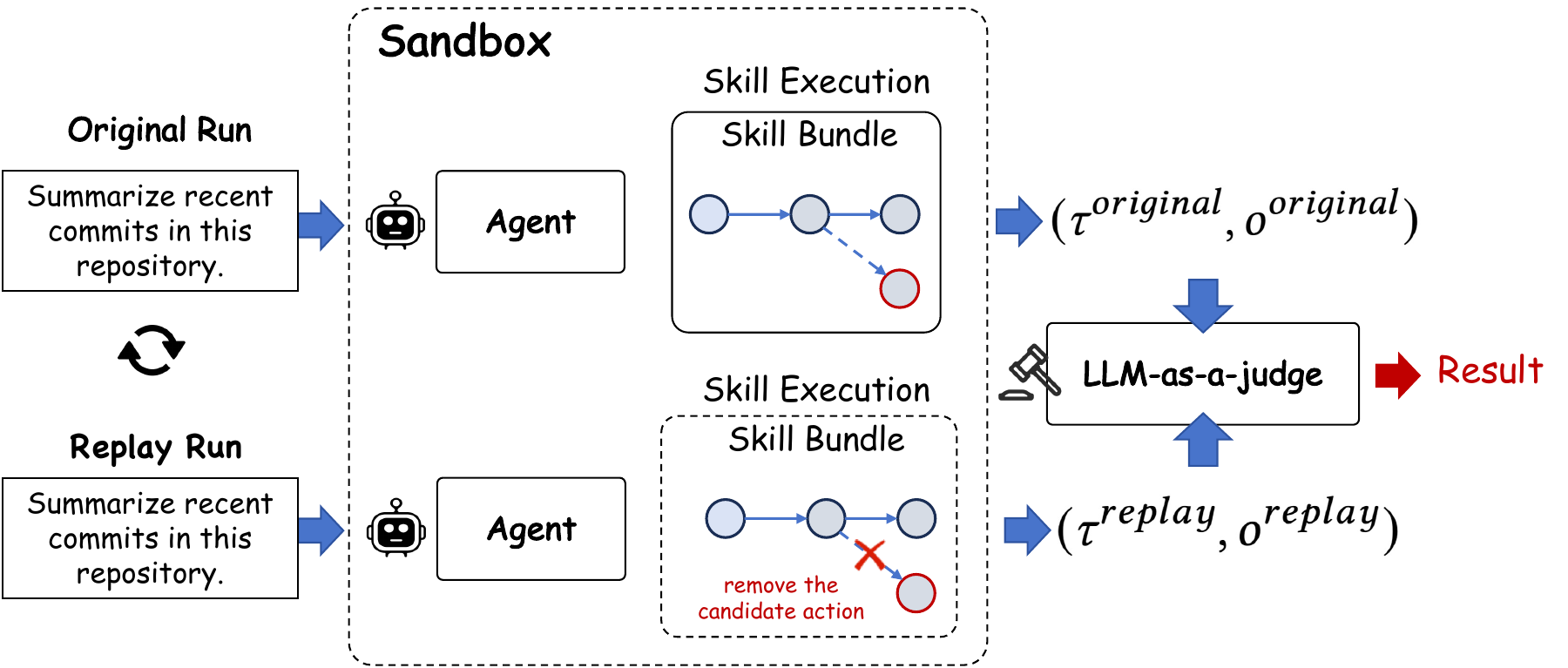}
    \Description{The same repository-summary request is executed twice in a sandbox: once with the original skill and once with a candidate action removed. Both runs produce an execution trace and a final output, which an LLM judge compares to assess whether removing the action preserves the task's required behavior and outcome.}
    \caption{The process of runtime task replay execution.}
    \label{fig:replay-based-confirmation}
\end{figure}

\phantomsection
\label{rev:runtime-task-validation}
\para{Runtime Task Replay Execution}
Following the definition in \Cref{subsec:problem statement}, \system{} validates action necessity through task replay execution. For a skill $\sigma$, a user prompt $p$, and a candidate action $a$, the replay run corresponds to the modified execution $\sigma \ominus a$, where $a$ is removed or neutralized while the remaining skill logic is preserved. The goal is to determine whether removing $a$ affects completion of the task requested by $p$. \hypertarget{rev:m1-runtime-alignment}{}\blue{The original run records the realized action instance used to evaluate $\mathsf{Auth}_{p}(a)$, while the original/replay pair supplies evidence for $\mathsf{CorePres}_{p}$ and $\mathsf{GoalSat}_{p}$.}

Specifically, for each instantiated user prompt and over-privilege candidate, \system{} executes the skill twice under the same task request, as illustrated in \Cref{fig:replay-based-confirmation}: \textbf{\textit{(1) the original run}}, which corresponds to $\tau(\sigma,p)$ and $o(\sigma,p)$, and \textbf{\textit{(2) the replay run}}, which corresponds to $\tau(\sigma \ominus a,p)$ and $o(\sigma \ominus a,p)$. For instruction-level candidates, \system{} removes the corresponding instruction step. For code-level candidates that are tightly coupled with surrounding logic, \system{} neutralizes only the candidate behavior, e.g., by replacing the target function body with an immediate return, to avoid artificial failures caused by broken dependencies.

\system{} builds an execution environment that isolates filesystem access, network communication, command execution, and other external side effects, so that the original and replay runs are conducted under controlled and comparable conditions. During each run, \system{} instruments the tested agent to record its execution trace and final user-visible output:
\[
(\tau^{original}, o^{original}) \quad \text{and} \quad (\tau^{replay}, o^{replay}),
\]
where $\tau^{original}=\tau(\sigma,p)$, $o^{original}=o(\sigma,p)$, $\tau^{replay}=\tau(\sigma \ominus a,p)$, and $o^{replay}=o(\sigma \ominus a,p)$.

\hypertarget{rev:m1-authorization-validation}{}\para{Task-Scope Authorization Check}
Using the original execution trace recorded above, \system{} evaluates $\mathsf{Auth}_{p}(a)$ only for candidates that are actually executed. In other words, \system{} determines whether the current action is authorized by the task (e.g., whether the user prompt explicitly requests access to bank-account data). A fixed-schema LLM scope judge compares the authority conveyed by the prompt with the realized action instance recorded in $\tau^{original}$, including its operation, data source, sink or destination, and side effects.

\para{Unnecessary Action Extraction}
Furthermore, \system{} determines whether a candidate action is unnecessary by checking the two conditions defined in \Cref{subsec:problem statement}: core task-flow preservation and replay goal satisfaction. Specifically, \system{} estimates the following two predicates using an LLM-as-a-judge approach, following prior work~\cite{llm-as-a-judge-1,llm-as-a-judge-2,llm-as-a-judge-3}. The judge receives the candidate and its provenance, the prompt, normalized original/replay traces, and their associated outputs. For core-flow preservation, \system{} normalizes both traces into ordered action summaries, including invoked actions, target objects, and summarized arguments, and checks whether the replay still preserves the core task flow and externally observable effects needed to satisfy the user intent after ablating the candidate. For goal satisfaction, \system{} uses the original execution as contextual evidence and checks whether the replay execution and its associated output still fulfill the user's requested task; neither the trace nor the output needs to be identical to its original counterpart.

\hypertarget{rev:m1-replay-criteria}{}\blue{For a candidate that is executed in the original run and removed or neutralized in the replay run, \system{} determines that the candidate is unnecessary under prompt $p$ when both replay predicates hold:}
{\color{black}\[
    \mathsf{CorePres}_{p}(\tau^{original},\tau^{replay})
    \ \wedge\
    \mathsf{GoalSat}_{p}(\tau^{original},\tau^{replay}).
\]}
\blue{When both predicates hold, they establish $\mathsf{Unnec}_{\sigma}(a,p)$. Otherwise, if the replay fails to preserve the prompt-required core flow or its result no longer satisfies the user's goal, the candidate is not confirmed as unnecessary for that task.}

\noindent \blue{\textbf{Over-privilege Action Confirmation.} \system{} then combines the replay-based necessity result with the task-scope authorization result. A candidate action executed in the original run is confirmed as over-privileged when $\neg\mathsf{Auth}_{p}(a)\vee\mathsf{Unnec}_{\sigma}(a,p)$.}

\subsection{Control-flow Privilege Constraining}

\begin{figure}[htbp]
    \centering
    \includegraphics[width=\linewidth]{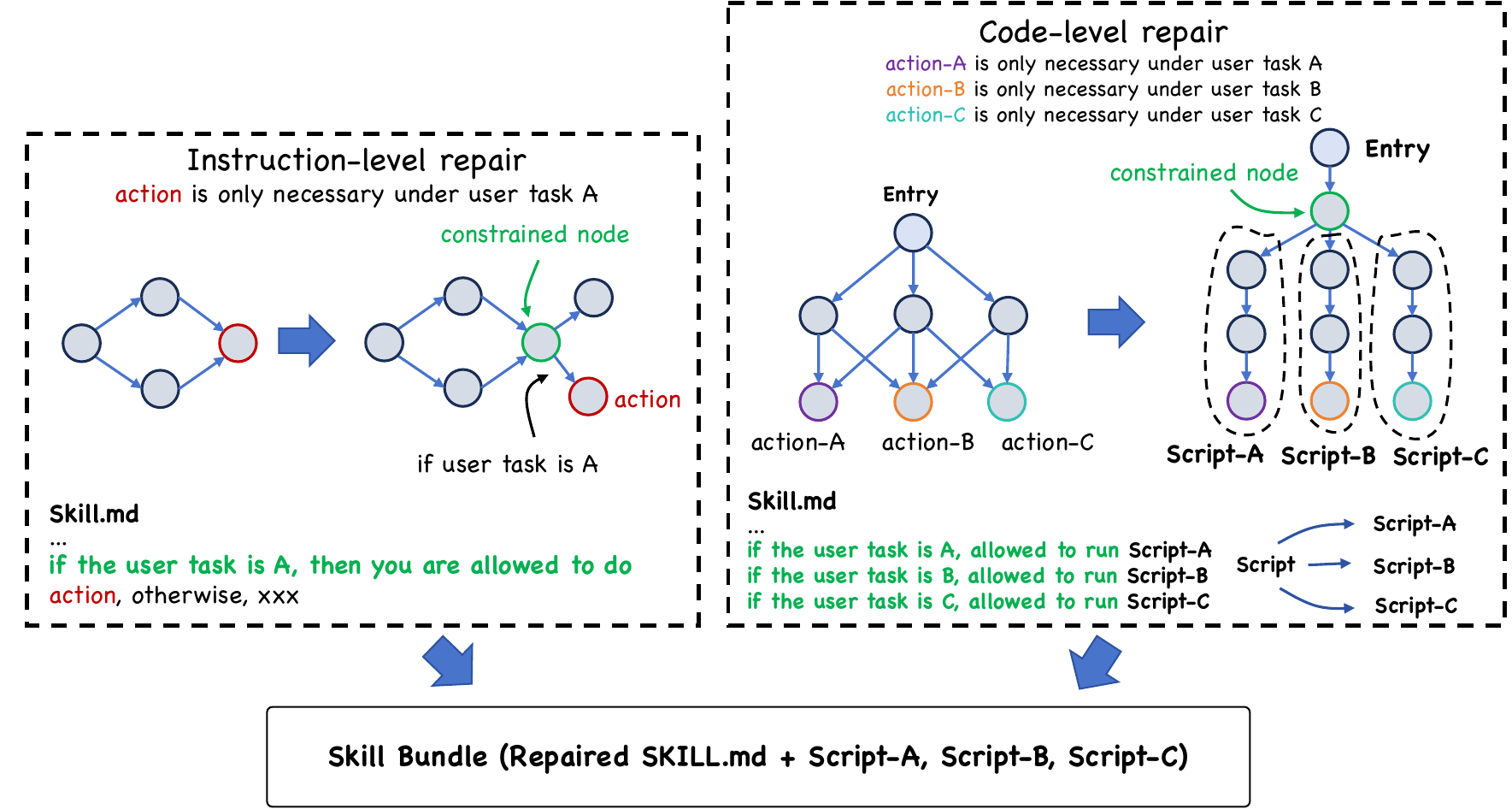}
    \Description{Instruction-level repair inserts a semantic guard that allows an action only when the user task requires it. Code-level repair reorganizes actions A, B, and C into separate scripts and adds task-conditioned routing instructions that select the corresponding script. Both transformations produce a repaired skill bundle.}
    \caption{Task-conditioned over-privilege control. Constraining validated over-privileged behaviors through semantic instruction constraints and task-specific code reorganization.}
    \label{fig:repair-process-2}
\end{figure}

The third module constrains validated over-privileged actions while preserving legitimate skill utility. Guided by validation results, \system{} applies localized control-flow transformations to the original \texttt{SKILL.md} and scripts. Instead of permanently deleting an action based only on a finite generated task set, \system{} removes the validated over-privileged action from the default execution path and makes it reachable only through an explicit task-conditioned control branch. In this way, over-privileged behaviors are no longer triggered implicitly, but can still be preserved behind semantic constraints if they are explicitly required by future user tasks. The constraining process is shown in \Cref{fig:repair-process-2}.

\para{Task-conditioned Constrained Node Generation}
Before rewriting the Skill artifacts, \system{} generates a task-conditioned constrained node for each validated over-privileged action. For an action $a=\langle \ell,op,obj\rangle$, \system{} uses the generated user tasks and their validation results to identify the task contexts under which $a$ should or should not be executed. Concretely, \system{} first converts each user task into a fixed-schema task-context descriptor, including the task intent, requested operation, target object, execution scope, any external destination, whether the user explicitly requests the side effect corresponding to $a$, and the validation result of $a$ under that task. It then normalizes semantically equivalent slots and groups descriptors into task-context clusters. For example, for $\textsf{send}(\textit{report}, \textit{Telegram})$, \system{} distinguishes tasks that only require local report generation from tasks that explicitly request sending or synchronizing the report to Telegram. The former should not satisfy the guard because they do not require external transmission.

Based on these clustered contexts, \system{} derives an allow condition $C_a$ for $a$. The condition is generated contrastively: it should allow contexts where the task explicitly requires the operation, object, scope, and side effect of $a$, while excluding contexts where $a$ was validated as over-privileged. \system{} then creates a constrained control node $v_a^c=\langle C_a,a\rangle$, where $C_a$ is the task-conditioned guard and $a$ is the guarded action. During execution, $v_a^c$ routes control to $a$ only when the current user task satisfies $C_a$; otherwise, the guarded action is skipped. We provide the detailed descriptor schema and a worked example of constrained node generation in Appendix~\ref{app:constrained-node-generation}.

\para{Guard Insertion and Script Reorganization}
For over-privileged instruction-level actions, \system{} inserts the constrained control node generated in the previous step before the corresponding action in the unified execution graph. The node is then projected back to \texttt{SKILL.md} as a semantic guard that specifies the task condition under which the action is allowed. \system{} keeps this guard at the instruction layer rather than converting it into a rigid keyword check, because semantically equivalent user tasks may differ in wording and should be routed by the agent's underlying LLM.

For over-privileged code-level behaviors, privilege constraining is more challenging because operations are often embedded in highly coupled execution logic rather than isolated statements. Directly deleting or guarding a code action may break data/control dependencies and damage the original functionality. Moreover, the task condition comes from natural language: semantically equivalent user requests may differ substantially in wording, while code-level guards must be expressed as rigid predicates (e.g., \texttt{if} conditions), causing semantic loss.

To address these challenges, \system{} adopts a two-level strategy. At the code level, it identifies the over-privilege branch and reorganizes the coupled script into task-specific execution units using LLM-assisted refactoring techniques~\cite{script-refactory-1,script-refactory-2,script-refactory-3}, such as extract-method-style refactoring and method-level restructuring. At the instruction level, \system{} adds a corresponding semantic constraint to \texttt{SKILL.md}, so that the agent's underlying LLM decides which script unit should be invoked under the current task semantics. In this way, high-dimensional task understanding remains at the instruction layer, while the code layer becomes modular and deterministic, with each separated script unit implementing only its corresponding task-specific behavior.

\para{Constraint Projection to Skill Artifacts}
After the constraints are determined on the unified execution graph, \system{} projects them back to the original skill artifacts: instruction-level constraints are written into \texttt{SKILL.md}, while code-level constraints are materialized as guarded branches or reorganized task-specific scripts. The final output is a least-privilege skill bundle containing the updated instruction file and constrained script set.

\section{Experiment}

In this section, we aim to answer the following questions:
\begin{itemize}[leftmargin=*]
    \item How effective is \system{} in detecting and validating over-privileged actions in terms of its end-to-end detection accuracy, the effectiveness of each module, and its performance compared with existing work?
    
    \item How prevalent are over-privileged skills in the current Agent Skill ecosystem?
    
    \item How effective is \system{} in constraining over-privileged skills? In particular, can the constrained skills preserve stable task utility while reducing over-privileged behaviors, and how do the original and constrained skills behave when deployed in real-world agents?
\end{itemize}

\subsection{Experimental Setup}
\label{subsec:experimental_setup}

\para{Dataset}
To evaluate \system{}, we construct two datasets for different purposes:
\begin{itemize}[leftmargin=*]
    \item \textbf{Large-scale measurement.} We collect Agent Skills from major public skill repositories and marketplaces, including Clawhub~\cite{clawhub} and Agent Skills Marketplace~\cite{skill-marketplace}. In total, we crawl 110,046 skills. After removing duplicates and unavailable bundles, we obtain 68,312 valid skill bundles for large-scale measurement of over-privileged behaviors in the wild.

    \item \textbf{Effectiveness and privilege-constraining evaluation.} From the above ecosystem-scale dataset, we manually sample and annotate 200 skills in total, including 100 skills with over-privileged actions and 100 benign skills without over-privileged actions. To reduce sampling bias in the manually labeled evaluation set, we sample Skills using a stratified strategy over marketplace source, Skill category, artifact type, and observable operation type. We provide the detailed dataset construction criteria in \Cref{app:dataset-construction}. Ground truth is assigned at the action-in-task level using the task-conditioned criterion in \Cref{subsec:problem statement}. Each Skill is independently inspected by two annotators, and cases without consensus are reviewed by a third annotator; if the available evidence remains insufficient, the action-in-task pair is conservatively treated as unconfirmed. This dataset is used for the quantitative evaluation of \system{}, as well as for the subsequent privilege-constraining evaluation.
\end{itemize}

\para{Implementation and Execution Environment}
All experiments are conducted on a Mac with an M4 Max processor and 36 GB of RAM. For task-replay validation and privilege-constraining validation, we build an execution environment that isolates filesystem access, network communication, and command execution. For LLM-dependent components in \system{}, we use GPT-5.1~\cite{gpt-5-1} as the default model. We adopt GPT-5.1 in our evaluation mainly because it offers a favorable cost-performance tradeoff, which is important for supporting our large-scale experiments while maintaining stable effectiveness across all stages (see \Cref{subsec:rq1} for details).

\subsection{Effectiveness of \system{}}
\label{subsec:rq1}

In this section, we evaluate the effectiveness of \system{} on the manually labeled 200-skill dataset introduced in \Cref{subsec:experimental_setup}. We focus on four aspects: the end-to-end effectiveness of \system{}, the effectiveness of over-privilege candidate extraction (\Cref{subsec:overreach-candidate-extraction}), the effectiveness of action over-privilege validation (\Cref{subsec:action-necessity-validation}), and the comparison against existing baselines.

\begin{table}[tbhp]
\centering
\caption{End-to-end effectiveness of \system{} for over-privileged skill detection.}
\label{tab:rq2_end2end}
\resizebox{\linewidth}{!}{%
\renewcommand{\arraystretch}{1.2}
\begin{tabular}{l|c|c|c|c|c}
\Xhline{0.8pt}
\multirow{2}{*}{\textbf{\system{}}}
 & \multirow{2}{*}{\textbf{TP/FP/TN/FN}}
& \textbf{Prec.}
& \textbf{Rec.}
& \textbf{F1}
& \textbf{Acc.} \\
&  &{\textbf{(\%)}}&{\textbf{(\%)}}&{\textbf{(\%)}}&{\textbf{(\%)}}\\
\hline
Skill-level      & 95/6/94/5       & 94.06 & 95.00 & 94.53 & 94.50 \\
Action-level     & 279/21/1,803/10  & 93.00 & 96.54 & 94.74 & 98.53 \\
Action-in-Task   & 472/33/197/35   & 93.47 & 93.10 & 93.28 & 90.77 \\
\Xhline{0.8pt}
\end{tabular}%
}
\end{table}

\para{End-to-End Effectiveness}
The overall end-to-end results are shown in \Cref{tab:rq2_end2end}. At the skill level, \system{} achieves 94.06\% precision, 95.00\% recall, and 94.53\% F1. At the action level, \system{} achieves 93.00\% precision, 96.54\% recall, and 94.74\% F1, indicating that it can further localize the specific actions responsible for over-privilege. At the action-in-task level, \system{} achieves 93.47\% precision, 93.10\% recall, and 93.28\% F1, showing that it can determine whether a concrete action is over-privileged under a specific user task. A small number of false positives and false negatives still remain, mainly due to the inherent instability of LLM-based agents, such as model hallucination and occasional failure to follow the intended instruction path during task execution. Despite these challenges, \system{} achieves consistently high effectiveness across all three levels.


\begin{table}[tbhp]
\centering
\caption{Cross-model effectiveness of over-privilege candidate extraction in \system{}.}
\label{tab:rq2_static}
\resizebox{\linewidth}{!}{%
\renewcommand{\arraystretch}{1.4}
\fontsize{16pt}{16pt}\selectfont
\begin{tabular}{l|c|c|c|c|c}
\Xhline{1.2pt}
\multirow{2}{*}{\textbf{Analysis LLM}} 
& \textbf{TP/FP/TN/FN} 
& \textbf{Prec.}
& \textbf{Rec.}
& \textbf{F1}
& \textbf{Acc.} \\
& \textbf{(\# Action Nodes)} & {\textbf{(\%)}} & {\textbf{(\%)}} & {\textbf{(\%)}} & {\textbf{(\%)}} \\
\hline
GPT-5.1                       & 281/129/1,695/8  & 68.54 & 97.23 & 80.31 & 93.52 \\ \hline
Gemini-3.1-Flash-Lite         & 247/203/1,621/42 & 54.89 & 85.47 & 66.86 & 88.40 \\
Claude-Sonnet-4.6             & 286/187/1,637/3  & 60.47 & 98.96 & 75.08 & 91.01 \\
DeepSeek-V3.2                 & 269/151/1,673/20 & 64.05 & 93.08 & 75.88 & 91.91 \\
\Xhline{1.2pt}
\end{tabular}%
}
\end{table}

\para{Over-Privilege Candidate Extraction (\Cref{subsec:overreach-candidate-extraction})}
The effectiveness of over-privilege candidate extraction is shown in \Cref{tab:rq2_static}. The goal of this module is to extract potentially problematic candidate behaviors from a large instruction-code search space. From the results, this stage is able to effectively retain nearly all true over-privileged behaviors, achieving 97.23\% recall at the action node level. These high-recall results show that the candidate extraction stage can reliably cover almost all truly over-privileged skills and actions before dynamic validation. Since this stage is built on the unified execution graph, we further manually evaluate the quality of the unified graph construction. The results show consistently high accuracy across nodes and edges. Detailed results are shown in Appendix~\ref{app:graph_quality}. Across instruction nodes and edges, code nodes and edges, and cross-layer links, the component-level F1 scores range from 92.62\% to 98.07\%.

For model selection in this module, the cross-model results show that GPT-5.1 provides the best overall balance between recall and precision. Claude-Sonnet-4.6 achieves the highest recall, but its stricter safety-oriented judgment introduces more false positives, which would increase the cost of downstream replay validation. Meanwhile, due to its intentionally conservative over-approximation design, this stage also introduces some false positives. However, such false positives are expected and are further filtered out by the subsequent dynamic validation stage. More importantly, by narrowing the original large execution space into a much smaller set of suspicious candidates, this module substantially reduces the unnecessary search space for the confirmation module.

\begin{table}[tbhp]
\centering
\caption{User-task instantiation with different prompt-generation LLMs.}
\label{tab:rq2_prompt_generation}
\resizebox{\linewidth}{!}{%
\renewcommand{\arraystretch}{1.5}
\fontsize{16pt}{14pt}\selectfont
\begin{tabular}{l|c|c|c}
\Xhline{1.2pt}
\raisebox{-2ex}{\rule{0pt}{5ex}}\textbf{LLM} & \makecell[c]{\textbf{\# Total}\\ \textbf{Chains}} & \makecell[c]{\textbf{\# Successfully}\\ \textbf{Triggered}}  & \textbf{Coverage (\%)} \\
\hline
GPT-5.1                       & 261 & 249 & 95.40 \\ \hline
Gemini-3.1-Flash-Lite         & \multirow{3}{*}{261} & 238 & 91.19 \\
Claude-Sonnet-4.6             &                      & 251 & 96.17 \\
DeepSeek-V3.2                 &                      & 244 & 93.49 \\
\Xhline{1.2pt}
\end{tabular}%
}
\end{table}

\para{Action Over-Privilege Validation (\Cref{subsec:action-necessity-validation})}
The effectiveness of action over-privilege validation is shown in \Cref{tab:rq2_prompt_generation,tab:rq2_dynamic_judge}. We evaluate whether the synthesized user tasks can effectively cover the task space represented by the extracted task chains of the skill. As shown in \Cref{tab:rq2_prompt_generation}, all evaluated prompt-generation LLMs achieve high coverage, indicating that this step is not tied to a single model family. Under the default GPT-5.1 setting, 249 out of 261 task chains can be successfully triggered, yielding a coverage of 95.40\%. Claude-Sonnet-4.6 achieves the highest coverage, but it also incurs the highest inference cost among the evaluated models. Considering both coverage and practical cost, we use GPT-5.1 as the default prompt-generation LLM. The high task-chain coverage shows that \textbf{\system{} can systematically instantiate user tasks from the graph-represented task space of a Skill, enabling broad task-conditioned over-privilege validation rather than relying on narrow case-by-case checking}.

\begin{table}[tbhp]
\centering
\caption{Over-privilege action confirmation with different judge LLMs.}
\label{tab:rq2_dynamic_judge}
\resizebox{\linewidth}{!}{%
\renewcommand{\arraystretch} {1.4}
\fontsize{16pt}{16pt}\selectfont
\begin{tabular}{l|c|c|c|c|c}
\Xhline{1.1pt}
\multirow{2}{*}{\textbf{Judge LLM}} 
& \textbf{TP/FP/TN/FN} 
& \textbf{Prec.}
& \textbf{Rec.}
& \textbf{F1}
& \textbf{Acc.} \\
& \textbf{(\# Action-in-Task Instances)} & {\textbf{(\%)}} & {\textbf{(\%)}} & {\textbf{(\%)}} &  {\textbf{(\%)}}\\
\hline
GPT-5.1                       & 472/33/197/35 & 93.47 & 93.10 & 93.28 & 90.77 \\ \hline
Gemini-3.1-Flash-Lite         & 428/70/160/79 & 85.94 & 84.42 & 85.17 & 79.78 \\
Claude-Sonnet-4.6             & 442/23/207/65 & 95.05 & 87.18 & 90.95 & 88.06 \\
DeepSeek-V3.2                 & 451/52/178/56 & 89.66 & 88.95 & 89.31 & 85.35 \\
\Xhline{1.1pt}
\end{tabular}%
}
\end{table}

We further evaluate the final over-privilege action confirmation, implemented using an LLM-as-a-judge approach. As shown in \Cref{tab:rq2_dynamic_judge}, GPT-5.1 achieves the best overall F1, with 93.47\% precision, 93.10\% recall, and 93.28\% F1, showing that \system{} can accurately determine whether a candidate action is truly over-privileged under the current task. Claude-Sonnet-4.6 achieves the highest precision, indicating that its confirmed over-privilege judgments are highly reliable. However, its lower recall suggests that it is stricter when comparing original and replay executions, and may reject some replay outputs that differ in wording or minor execution details even when the core task is preserved. Therefore, although Claude-Sonnet-4.6 is more conservative, GPT-5.1 provides a better overall balance for replay-based validation.

\begin{table}[tbhp]
\centering
\caption{Effectiveness of \system{} compared with that of existing methods.}
\label{tab:rq2_baseline_skill_est}
\resizebox{\linewidth}{!}{%
\renewcommand{\arraystretch}{1.35}
\fontsize{14pt}{14pt}\selectfont
\begin{tabular}{l|c|c|c|c}
\Xhline{1.1pt}
\textbf{Method} & \textbf{TP/FP/TN/FN} & \textbf{Prec. (\%)} & \textbf{Rec. (\%)} & \textbf{F1 (\%)} \\
\hline
\textit{Skill Scanner}~\cite{skill-scanner}    & 81/26/74/19 & 75.70 & 81.00 & 78.26 \\
\textit{Caterpillar}~\cite{caterpillar}      & 62/41/59/38 & 60.19 & 62.00 & 61.08 \\
\textit{MASB}~\cite{masb}             & 19/28/72/81 & 40.43 & 19.00 & 25.85 \\
\textit{Skill-Sec-Scan}~\cite{skill-sec-scan}   & 45/37/63/55 & 54.88 & 45.00 & 49.45 \\
\textit{Nova-Proximity}~\cite{nova-pro}   & 14/32/68/86 & 30.43 & 14.00 & 19.18 \\
\textit{MalSkills}~\cite{malskill}        & 86/12/88/14 & 87.76 & 86.00 & 86.87 \\
\hline
\textbf{\system{}}                   & \textbf{95/6/94/5} & \textbf{94.06} & \textbf{95.00} & \textbf{94.53} \\
\Xhline{1.1pt}
\end{tabular}%
}
\end{table}

\para{Comparison with Baselines}
We compare \system{} against six representative skill detection tools covering static analysis, LLM-based analysis, and dynamic analysis: \textit{MalSkills}~\cite{malskill}, \textit{Skill Scanner}~\cite{skill-scanner}, \textit{Nova-Proximity}~\cite{nova-pro}, \textit{Skill-Sec-Scan}~\cite{skill-sec-scan}, \textit{Caterpillar}~\cite{caterpillar}, and \textit{MASB}~\cite{masb}. To the best of our knowledge, these constitute the most representative publicly available baselines for end-to-end skill detection in the current ecosystem. For LLM-based baselines, we use the same underlying model, GPT-5.1, for a fair comparison. As shown in \Cref{tab:rq2_baseline_skill_est}, \system{} achieves the best skill-level detection performance. This result shows that prior tools, which mainly provide coarse-grained skill-level judgments, are insufficient for task-conditioned over-privilege detection because they may miss many fine-grained over-privilege instances (i.e., cases in which an action becomes over-privileged under a concrete user task).

\begin{table}[tbhp]
\centering
\caption{Ablation study of key design choices in \system{} for over-privilege detection.}
\label{tab:rq2_ablation}
\resizebox{\linewidth}{!}{%
\renewcommand{\arraystretch}{1.25}
\begin{tabular}{l|c|c|c|c}
\Xhline{0.8pt}
\textbf{Variant}
& \textbf{TP/FP/TN/FN}
& \textbf{Prec. (\%)}
& \textbf{Rec. (\%)}
& \textbf{F1 (\%)} \\
\hline
\makecell[l]{\textit{w/o Unified Graph}}
& 73/25/75/27
& 74.49
& 73.00
& 73.74 \\ \hline

\makecell[l]{\textit{w/o Graph-guided}\\ \textit{Task Generation}}
& 76/9/91/24
& 89.41
& 76.00
& 82.16 \\ \hline

\makecell[l]{\textit{w/o Dynamic Validation}}
& 82/25/75/18
& 76.64
& 82.00
& 79.22 \\ \hline

\textbf{\system{}}
& \textbf{95/6/94/5}
& \textbf{94.06}
& \textbf{95.00}
& \textbf{94.53} \\
\Xhline{0.8pt}
\end{tabular}%
}
\end{table}

\para{Ablation Study}
To further understand the contribution of the key detection designs in \system{}, we compare the full detection pipeline with three simplified variants on the same 200-skill dataset. Since these modules provide required intermediate representations for downstream stages, directly removing one module would break the end-to-end pipeline. We therefore use replacement ablations that replace each key design with an executable alternative while keeping the final skill-level detection task unchanged. All LLM-based components in the variants use GPT-5.1, matching the full \system{} setting.

The first variant, \textit{w/o Unified Graph}, replaces graph-based action modeling with direct LLM analysis over raw skill artifacts. The second variant, \textit{w/o Graph-guided Task Generation}, keeps candidate extraction and replay validation, but replaces graph-chain-based task generation with skill profile-only prompt generation under the same prompt budget. The third variant, \textit{w/o Dynamic Validation}, replaces original/replay execution with a static LLM necessity judge that does not observe runtime traces or outputs.

The results are shown in \Cref{tab:rq2_ablation}. Without the unified graph, F1 drops to 73.74\%, because raw artifact-level reasoning lacks explicit action boundaries, dependency context, and provenance. Without graph-guided task generation, F1 drops to 82.16\%; this variant still maintains relatively high precision, but its recall decreases because profile-only prompts miss some candidate-reachable task paths. Without dynamic validation, F1 drops to 79.22\%, as the static judge lacks original/replay trace and output evidence, leading to both false positives and false negatives. These results confirm that graph-based action modeling, graph-guided task coverage, and dynamic over-privilege validation provide complementary benefits.

\begin{table}[tbhp]
\color{black}
\centering
\caption{Summary of token cost on the 200-skill dataset.}
\label{tab:token_cost_summary}
\resizebox{\columnwidth}{!}{%
\renewcommand{\arraystretch}{1.08}
\fontsize{8pt}{9pt}\selectfont
\begin{tabular}{l|r|r|r}
\Xhline{0.7pt}
\textbf{Module / Stage}
& \makecell[c]{\textbf{\# LLM}\\\textbf{Calls}}
& \makecell[c]{\textbf{Total Tokens}\\\textbf{(M)}}
& \makecell[c]{\textbf{Avg. Tokens}\\\textbf{/ Skill (M)}} \\
\hline
Unified Graph Construction
& 200
& 0.469
& 0.00235 \\
Action Consistency Screening
& 2,113
& 2.387
& 0.01194 \\
User Task/Prompt Instantiation
& 261
& 0.229
& 0.00115 \\
Over-Privilege Validation
& 2,211
& 4.389
& 0.02195 \\
Privilege Constraining
& 666
& 1.862
& 0.00931 \\ \hline
\textbf{Overall}
& \textbf{5,451}
& \textbf{9.337}
& \textbf{0.04669} \\
\Xhline{0.7pt}
\end{tabular}%
}
\end{table}

\noindent \hypertarget{rev:m2-token-cost}{}\blue{\textbf{Token Cost.} We further measure the token cost of \system{} across detection and privilege constraining. The stage-wise summary is reported in \Cref{tab:token_cost_summary}, with detailed call and token accounting provided in Appendix~\ref{app:token_cost}.}

\blue{Across 200 Skills, \system{} issues 5,451 LLM calls and consumes 9.337M tokens (8.292M input and 1.045M output), averaging 46.69K tokens per Skill. Over-privilege validation is the largest source of total token use (4.389M), while privilege constraining has the highest per-unit cost (6.67K per repaired action). Note that the analysis is performed once by a marketplace or enterprise when a Skill version is published or admitted; its results and intermediate artifacts can then be cached and reused. Further savings are possible by selectively using lower-cost models. For several modules (e.g., candidate extraction and user task instantiation), the gap between models is small because the framework design and low task complexity reduce the demand on the model. As a result, \system{} can use cheaper models (e.g., DeepSeek-V3.2) for these tasks and still obtain comparable results.}


\begin{table*}[t]
\centering
\caption{Overall statistics of over-privilege candidates and validated over-privilege actions at the skill, unified-graph, and user-task levels.}
\label{tab:rq1_overall}
\resizebox{\textwidth}{!}{%
\renewcommand{\arraystretch}{1}
\fontsize{8pt}{10pt}\selectfont
\begin{tabular}{l|c|c|c|c|c|c}
\Xhline{0.7pt}
\multirow{3}{*}{\textbf{Source}}
& \multicolumn{2}{c|}{\textbf{Skill Level}}
& \multicolumn{2}{c|}{\textbf{Action Level}}
& \multicolumn{2}{c}{\textbf{User Task/Prompt Level}} \\
\cline{2-7}
& \rule{0pt}{4ex}\makecell[c]{\# Over-privilege\\ Candidates}
& \makecell[c]{\# Involving\\ Over-privilege}
& \makecell[c]{\# Total\\ (Action Nodes)}
& \makecell[c]{\# Over-privilege Candidates\\ (Action Nodes)}
& \makecell[c]{\# Total\\ User Tasks}
& \makecell[c]{\# Involving\\ Over-privilege} \\
\hline
\rule{0pt}{2ex}Instruction-level & 6,997 & 5,541 & 63,274 & 9,846 & \multirow{3}{*}{\textbf{16,743}} & 6,075 \\
Code-level        & 1,891 & 1,498 & 22,847 & 3,171 &                         & 1,642 \\
Both levels       & 560   & 449   & 8,931  & 1,429 &                         & 684 \\ \hline
\textbf{Overall}  & \textbf{8,328} & \textbf{6,590} & \textbf{77,190} & \textbf{11,588} & \textbf{16,743} & \textbf{7,033} \\
\Xhline{0.7pt}
\end{tabular}%

}
\vspace{1mm}
\parbox{\textwidth}{\footnotesize\textit{Note.} The instruction-level and code-level rows are non-exclusive. ``Both levels'' reports their intersection, while ``Overall'' reports deduplicated counts.}
\end{table*}

\begin{tcolorbox}[
    colback=gray!10,
    colframe=gray!50,
    arc=0pt,
    outer arc=0pt,
    boxrule=0.3pt,
    left=2pt, right=2pt, top=0pt, bottom=0pt
]

\para{\textit{Takeaway-1}}
\system{} achieves effective end-to-end over-privilege detection through candidate extraction and over-privilege validation. Compared with existing baselines, \system{} achieves fine-grained over-privileged action detection, which provides the necessary foundation for subsequent privilege constraining.
\end{tcolorbox}

\subsection{Over-Privileged Skills in the Wild}
\label{subsec:rq2}

In this section, we aim to measure how over-privilege manifests in real-world Agent Skills.

\para{Ecosystem-Scale Prevalence}
The overall measurement results are shown in \Cref{tab:rq1_overall}. Among the 68,312 valid skills, 8,328 skills contain at least one over-privilege candidate, and \system{} confirms that 6,590 skills exhibit validated over-privilege. This shows that over-privileged actions occur on a substantial scale in today's Agent Skill ecosystem.

At the unified-graph level, \system{} extracts 77,190 unique action nodes in total, among which 11,588 nodes are identified as over-privilege candidates. At the user-task level, 7,033 of the 16,743 instantiated user tasks are confirmed to trigger over-privileged behaviors, demonstrating that the problem is practically triggerable and frequently arises under realistic user requests. To calibrate these large-scale claims, we further conduct a post-hoc manual validation of skill-level positives, graph-level candidate nodes, and user-task-level positives detected by \system{}. The results confirm weighted validity rates of 94.16\%, 92.97\%, and 93.53\%, respectively, with the detailed sampling protocol and confidence intervals reported in Appendix~\ref{app:human_audit} and \Cref{tab:posthoc_human_audit}. The source-stratified audit covers 300 skill-level positives, 200 graph-level candidate nodes, and 200 user-task-level positives; code-involving cases are oversampled and the reported rates are population-weighted. These high validity rates indicate that the large-scale findings reported by \system{} are stable and largely reflect real over-privileged behaviors.

\Cref{tab:rq1_overall} shows that instruction-level cases dominate the current ecosystem. Specifically, 6,997 skills involve instruction-level candidates, and 5,541 skills are ultimately validated as instruction-level over-privilege. In comparison, 1,891 skills involve code-level candidates and 1,498 skills are validated after dynamic checking. These source-level counts are non-exclusive: 560 candidate skills and 449 validated skills exhibit behaviors induced by both instructions and code, while the overall counts are deduplicated. This result indicates that the natural-language instruction layer is currently the primary carrier of over-privileged behaviors. Since files such as \texttt{SKILL.md} directly govern high-level task procedures and execution policies, unnecessary or over-broad instructions can easily push the agent beyond the user's intended task boundary. Meanwhile, the non-trivial number of code-level cases also shows that bundled scripts remain an important channel for hidden behaviors.

\begin{table}[tbhp]
\centering
\caption{Breakdown of validated over-privilege actions by pattern.}
\label{tab:rq1_pattern}
\resizebox{\columnwidth}{!}{%
\renewcommand{\arraystretch}{1.5}
\fontsize{14pt}{14pt}\selectfont
\begin{tabular}{l|c|c|c|c}
\Xhline{1.5pt}
\textbf{Pattern}
& \rule{0pt}{3.5ex}\makecell[c]{Skill\\ Level}
& \makecell[c]{Skill\\ Level Ratio (\%)}
& \makecell[c]{User Task\\ Level} 
& \makecell[c]{User Task\\ Level Ratio (\%)} \\
\hline
\rule{0pt}{2.2ex}Over-collection & 2,961 & 44.93 & 3,158 & 44.90 \\
Over-execution  & 3,629 & 55.07 & 3,875 & 55.10 \\
\Xhline{1.5pt}
\end{tabular}%
}
\end{table}

\para{Distribution of Over-Privilege Patterns}
To further understand the nature of over-privilege in the wild, we next analyze which kinds of actions dominate the observed over-privileged behaviors. To this end, we group validated over-privileged behaviors into two patterns according to their functional effect. The first is \emph{over-collection}, which includes sensitive data access, excessive information collection, and data transmission or external exposure beyond the user’s intended task. The second is \emph{over-execution}, which includes high-impact operations such as SQL execution, code execution, command execution, and other actions that invoke extra capabilities outside the current task boundary.

The pattern-level results are shown in \Cref{tab:rq1_pattern}. We find that both over-collection and over-execution are common in real-world skills, but over-execution is slightly more prevalent. Specifically, over-execution affects 3,629 skills (55.07\%) and 3,875 user tasks (55.10\%), while over-collection affects 2,961 skills (44.93\%) and 3,158 user tasks (44.90\%). This result suggests that the dominant least-privilege failures in today’s Skill ecosystem combine data access and action execution that are beyond user intent. Many skills exceed user intent not only by collecting information beyond task needs but also by actually invoking extra operations or capabilities that are not required for completing the requested task.

Additionally, we observe that validated over-privileged behaviors are concentrated in capability-rich categories, especially tools (2,211 skills, 33.55\%), productivity-tools (1,608 skills, 24.40\%), and llm-ai (750 skills, 11.38\%). The detailed category breakdown is provided in Appendix~\ref{app:rq2_details}.

\begin{tcolorbox}[
    colback=gray!10,
    colframe=gray!50,
    arc=0pt,
    outer arc=0pt,
    boxrule=0.3pt,
    left=2pt, right=2pt, top=0pt, bottom=0pt
]

\para{\textit{Takeaway-2}}
Overall, our measurement shows that over-privilege is widespread in real-world Agent Skills, suggesting that today's third-party Skill ecosystem suffers from a systematic least-privilege boundary problem.
\end{tcolorbox}

\subsection{Effectiveness of Privilege Constraining}
\label{subsec:rq3}

In this section, we evaluate whether \system{} can constrain the over-privileged actions detected in \Cref{subsec:rq1} (\Cref{tab:rq2_end2end}, \Cref{subsec:rq1}). We measure overall suppression, effectiveness on real-world agents, and utility preservation.

\begin{table}[tbhp]
\centering
\caption{Overall effectiveness of \system{} in constraining over-privileged skills.}
\label{tab:rq3_overall_repair}
\resizebox{\linewidth}{!}{%
\renewcommand{\arraystretch}{1.1}
\begin{tabular}{l|c|c|c}
\Xhline{0.8pt}
\textbf{Metric} & \textbf{Orig.} & \textbf{Constr.} & \textbf{Reduct.} \\
\hline
\# Skills w/ over-privilege & 95 & 13 & \textbf{-82} \\
Over-privileged Skill rate (\%) & 100.00 & 13.68 & \textbf{-86.32} \\
\# Actions w/ over-privilege & 279 & 31 & \textbf{-248} \\
\# Action-in-task instances & 472 & 54 & \textbf{-418} \\
\# Task chains w/ over-privilege & 249 & 27 & \textbf{-222} \\ \hline
\textbf{Action-in-task suppression rate} & \multicolumn{3}{c}{\textbf{88.56\%}} \\
\Xhline{0.8pt}
\end{tabular}%
}
\footnotesize{\textit{Note:} Orig., Constr., and Reduct.\ denote Original, Constrained, and Reduction, respectively; w/ denotes with.}
\end{table}

\para{Overall Constraining Effectiveness}
The overall constraining results are shown in \Cref{tab:rq3_overall_repair}. After privilege constraining, the number of skills triggering over-privileged behaviors decreases from 95 to 13, corresponding to an 86.32\% reduction. At the action level, the number of triggered over-privileged actions is reduced from 279 to 31, while the number of triggered action-in-task instances decreases from 472 to 54. Meanwhile, the number of task chains triggering over-privileged behaviors drops from 249 to 27. Overall, \system{} achieves an 88.56\% action-in-task suppression rate. These results show that the constrained skills can substantially suppress over-privileged behaviors at the finer-grained action and task levels.

Although the constraining process is highly effective overall, it does not completely eliminate all over-privileged behaviors. The remaining cases mainly come from two sources: in some skills, the added constraints are diluted by surrounding complex context, causing the agent to miss the intended restriction during execution; in other cases, the current LLM-based program separation still cannot perfectly isolate all task-specific over-privileged logic. Nevertheless, the overall results already show that \system{} can effectively convert detected over-privileged skills into substantially safer least-privilege versions. Moreover, the additional token cost incurred by privilege constraining is included in the stage-wise accounting in \Cref{tab:token_cost_summary}, with further analysis in Appendix~\ref{app:token_cost}.

\begin{table}[tbhp]
\centering
\caption{Effectiveness of constrained skills on real-world agents.}
\label{tab:rq3_agent_repair}
\resizebox{\linewidth}{!}{%
\renewcommand{\arraystretch}{1.6}
\fontsize{20pt}{20pt}\selectfont
\begin{tabular}{l|c|c|c|c}
\Xhline{2.0pt}
\multirow{2}{*}{\textbf{Agent}}
& \multicolumn{2}{c|}{\makecell[c]{\textbf{Actions w/ Over-privilege}}}
& \multirow{2}{*}{\textbf{\# Reduced}}
& \multirow{2}{*}{\makecell[c]{\textbf{Reduction} \\ \textbf{Rate} (\%)}} \\
\cline{2-3}
& \textbf{\# Original} & \textbf{\# Constrained} & & \\
\hline
Claude Code & 279 & 34 & \textbf{-245} & \textbf{87.81} \\
Codex       & 279 & 29 & \textbf{-250} & \textbf{89.61} \\
Cursor      & 279 & 38 & \textbf{-241} & \textbf{86.38} \\
Trae        & 279 & 27 & \textbf{-252} & \textbf{90.32} \\
\Xhline{2.0pt}
\end{tabular}%
}
\end{table}

\para{Constraining Effectiveness on Real-World Agents}
We further evaluate whether the constrained skills remain effective when deployed on representative real-world agents, and the results are shown in \Cref{tab:rq3_agent_repair}. To this end, we select four widely used real-world agent products from different companies, including Claude Code~\cite{claude-code}, Codex~\cite{codex}, Cursor~\cite{cursor}, and Trae~\cite{trae}, as representative deployment targets. For fairness, all agents are evaluated using the same 95 skills from \Cref{tab:rq3_overall_repair} and the same user prompt set. Detailed deployment settings, including evaluation inputs and permission settings, are provided in Appendix~\ref{app:agent_setup}. The generated constrained bundles are evaluated without manual editing. All runtimes use an isolated workspace and a consistent non-interactive policy that pre-authorizes only the sandboxed operations required by the test; this platform permission is distinct from task-scope authorization. Across all these agents, the number of triggered over-privileged actions is consistently and substantially reduced after privilege constraining.

These results show that the privilege constraints produced by \system{} remain effective across multiple real-world agent runtimes. The constrained skills are practically deployable: once an over-privileged skill is constrained by \system{}, the reduction in over-privileged behaviors can be observed when the same skill is executed by different mainstream agent systems.


\begin{table}[t]
\centering
\caption{Utility preservation of constrained skills on legitimate tasks.}
\label{tab:rq3_utility_baseline}
\resizebox{\linewidth}{!}{%
\renewcommand{\arraystretch}{1.25}
\setlength{\tabcolsep}{4pt}
\begin{tabular}{l|c|c|c}
\Xhline{1.1pt}
\textbf{Strategy}
& \makecell[c]{\textbf{Task}\\\textbf{Completion}}
& \makecell[c]{\textbf{Output}\\\textbf{Eq.}}
& \makecell[c]{\textbf{Core Trace}\\\textbf{Eq.}} \\
\hline
\makecell[c]{Direct Removal (Baseline)}
& \makecell[c]{144/249\\(57.83\%)}
& \makecell[c]{130/249\\(52.21\%)}
& \makecell[c]{112/249\\(44.98\%)} \\
\hline
\makecell[l]{\textbf{\system{} Constraining}}
& \makecell[c]{\textbf{249/249}\\\textbf{(100.00\%)}}
& \makecell[c]{\textbf{236/249}\\\textbf{(94.78\%)}}
& \makecell[c]{\textbf{232/249}\\\textbf{(93.17\%)}} \\
\Xhline{1.1pt}
\end{tabular}}
\end{table}

\para{Utility Preservation}
Finally, we evaluate whether constrained skills preserve legitimate functionality, as shown in \Cref{tab:rq3_utility_baseline}. We compare \system{} with a direct-removal baseline, which removes or neutralizes the validated over-privileged actions without adding task-conditioned control constraints. This baseline represents a straightforward mitigation strategy.

The results show that direct removal substantially degrades legitimate task utility. Among the 249 legitimate tasks, only 144 tasks can still be completed (57.83\%), 130 tasks preserve output equivalence (52.21\%), and 112 tasks preserve the core execution trace (44.98\%). In contrast, \system{} preserves task completion for all 249 tasks after privilege constraining. It also preserves output equivalence for 236 tasks (94.78\%) and core execution trace equivalence for 232 tasks (93.17\%). These results show that the utility benefit of \system{} comes from task-conditioned control rather than unconditional deletion: over-privileged actions are removed from the default execution path, but remain reachable when the current task authorizes and requires them.

A small amount of utility loss remains. In some borderline tasks, the inserted constraints make the agent more cautious, occasionally leading to conservative execution or slight changes in task-relevant behavior. Even so, compared with direct removal, \system{} preserves substantially more legitimate functionality while still suppressing over-privileged behavior. This supports the need for fine-grained, task-conditioned privilege constraining in Agent Skills.

\begin{tcolorbox}[
    colback=gray!10,
    colframe=gray!50,
    arc=0pt,
    outer arc=0pt,
    boxrule=0.3pt,
    left=2pt, right=2pt, top=0pt, bottom=0pt
]

\para{\textit{Takeaway-3}}
\system{} can effectively constrain over-privileged Agent Skills by substantially suppressing fine-grained over-privileged behaviors, while remaining effective on representative real-world agents and preserving the vast majority of legitimate task utility.
\end{tcolorbox}

\section{Discussion}

Our study shows that over-privilege is prevalent in current Agent Skills and is fundamentally task-conditioned. The same instruction- or code-level action may be acceptable under one task but over-privileged under another, making coarse-grained benign/malicious skill labels insufficient. Effective Skill governance therefore requires fine-grained analysis that reasons over concrete user tasks and identifies which specific action exceeds the intended task boundary.

\para{Skill Design Guidelines}
Our findings suggest that future Skill design should follow the principles of \emph{data minimization} and \emph{least privilege}. Skills should avoid vague instructions that call for collecting ``sufficient context'' or causing implicit side effects beyond the user's request. Instead, each functional module should be organized as an atomic operation with a clear task boundary, so that sensitive behaviors such as external transmission or sensitive-data access are invoked only when explicitly required. This design reduces unnecessary risk and makes later detection, diagnosis, and privilege constraining easier.

\para{Limitations}
\label{sec:limitations}
Our approach has several limitations. First, LLM-based agent execution remains unstable, and nondeterministic run-to-run variation may affect replay consistency and final judgments. Second, code-level privilege constraining depends on LLM-assisted program separation, which may be imperfect for highly coupled scripts. Third, for Skills requiring external resource fixtures, our current task generation relies on ad hoc integration of several fixture-generation tools, which may be difficult to extend to complex domain-specific inputs or authenticated external services. Future work can develop more systematic resource-specification extraction and extensible fixture synthesis for resource-dependent Skill validation. These limitations motivate future work on more robust task-space exploration, more stable replay environments, and stronger program restructuring for Agent Skill least-privilege enforcement.

\section{Related Work}

\para{Privacy Compliance}
Privacy compliance studies whether software systems collect, process, and transmit data within a necessary and proportionate scope~\cite{data-minimization-1,data-minimization-2,gdpr,saltzer1975protection,nist80053}, often through privacy policies, permission systems, information-flow control, and data-leakage analysis~\cite{privacy-policy-analysis,felt2011android,information-flow,privacy-leakage}. In LLM agents, compliance becomes dynamic because agents decide which tools, files, and data flows are used during task execution~\cite{agentdojo,AgentRaft,shi2025progent}. Agent Skills further expose this issue through both \texttt{SKILL.md} instructions and bundled scripts. Our work operationalizes data minimization and least privilege at the action level by checking whether each action is authorized and necessary for the current user task.

\para{LLM Agent Security}
Recent studies examine prompt injection, tool misuse, unsafe code or command execution, data exfiltration through tool chains, and attacks against GUI, web, database, and coding agents~\cite{agentdojo,Agentfuzz,AgentRaft,ChainFuzzer,MASTERKEY,promptinjectionssqlinjection,trojansql,Demystifying,ASB,philosopher,promptinject}. Existing defenses constrain agent behavior through sandboxing, prompt isolation, policy enforcement, or runtime monitoring~\cite{liu2024formalizingbenchmarkingpromptinjection,kim2025prompt,SecAlign,inject_defend_1,inject_defend_2,cStruQ,suo2024signedpromptnewapproachprevent,isolated}. These works address broad agent threats, while \system{} focuses on task-conditioned least-privilege violations inside reusable Skill bundles.

\para{Agent Skill Security}
Recent work and tools study Skill-specific risks such as malicious instructions, hidden code execution, credential leakage, reverse shells, prompt injection, and cross-artifact attacks~\cite{schmotz2025agent,malskill,skillprobe,skillattack,skillscan,masb,agent-skill-marketplace-security,schmotz2026skill,jia2026skillject}. Most determine whether a Skill is globally malicious by scanning metadata, instructions, scripts, or runtime behavior. In contrast, \system{} targets functionally plausible Skills whose actions cross the task-authorization or necessity boundary under specific user tasks, and provides necessity-aware replay validation plus utility-preserving privilege constraining.

\section{Conclusion}
\label{sec:conclusion}
This paper presents \system{}, a framework for fine-grained least-privilege enforcement in Agent Skills. \system{} combines unified execution graph construction, over-privilege candidate extraction, action over-privilege validation, and control-flow privilege constraining to detect and constrain actions that exceed the user's current task scope. In the effectiveness evaluation, \system{} achieves 94.53\% skill-level F1. At ecosystem scale, \system{} validates 6,590 over-privileged Skills among 68,312 real-world Skills, showing that least-privilege violations are prevalent. After applying the constraints, \system{} reduces triggered action-in-task instances by 88.56\% while preserving legitimate task completion.

\section*{Artifacts}
The SkillScope source code and prompts are available at \url{https://github.com/Jiangrong-W/SkillScope_CCS2026}.


\bibliographystyle{ACM-Reference-Format}
\bibliography{bib-full}

@article{schmotz2025agent,
  title={Agent Skills Enable a New Class of Realistic and Trivially Simple Prompt Injections},
  author={Schmotz, David and Abdelnabi, Sahar and Andriushchenko, Maksym},
  journal={arXiv preprint arXiv:2510.26328},
  year={2025}
}

@inproceedings{Demystifying,
  title={Demystifying rce vulnerabilities in llm-integrated apps},
  author={Liu, Tong and Deng, Zizhuang and Meng, Guozhu and Li, Yuekang and Chen, Kai},
  booktitle={Proceedings of the 2024 on ACM SIGSAC Conference on Computer and Communications Security},
  pages={1716--1730},
  year={2024}
}

@article{isolated,
  title={Drift: Dynamic rule-based defense with injection isolation for securing llm agents},
  author={Li, Hao and Liu, Xiaogeng and Chiu, Hung-Chun and Li, Dianqi and Zhang, Ning and Xiao, Chaowei},
  journal={arXiv preprint arXiv:2506.12104},
  year={2025}
}

@inproceedings{suo2024signedpromptnewapproachprevent,
  title={Signed-prompt: A new approach to prevent prompt injection attacks against llm-integrated applications},
  author={Suo, Xuchen},
  booktitle={AIP Conference Proceedings},
  volume={3194},
  number={1},
  pages={040013},
  year={2024},
  organization={AIP Publishing LLC}
}

@misc{inject_defend_1,
      title={Defense against Prompt Injection Attacks via Mixture of Encodings}, 
      author={Ruiyi Zhang and David Sullivan and Kyle Jackson and Pengtao Xie and Mei Chen},
      year={2025},
      eprint={2504.07467},
      archivePrefix={arXiv},
      primaryClass={cs.CL},
      url={https://arxiv.org/abs/2504.07467}, 
}

@misc{inject_defend_2,
      title={Encrypted Prompt: Securing LLM Applications Against Unauthorized Actions}, 
      author={Shih-Han Chan},
      year={2025},
      eprint={2503.23250},
      archivePrefix={arXiv},
      primaryClass={cs.CR},
      url={https://arxiv.org/abs/2503.23250}, 
}

@inproceedings{cStruQ,
  title={$\{$StruQ$\}$: Defending against prompt injection with structured queries},
  author={Chen, Sizhe and Piet, Julien and Sitawarin, Chawin and Wagner, David},
  booktitle={34th USENIX Security Symposium (USENIX Security 25)},
  pages={2383--2400},
  year={2025}
}

@inproceedings{SecAlign,
  title={Secalign: Defending against prompt injection with preference optimization},
  author={Chen, Sizhe and Zharmagambetov, Arman and Mahloujifar, Saeed and Chaudhuri, Kamalika and Wagner, David and Guo, Chuan},
  booktitle={Proceedings of the 2025 ACM SIGSAC Conference on Computer and Communications Security},
  pages={2833--2847},
  year={2025}
}

@inproceedings{liu2024formalizingbenchmarkingpromptinjection,
  title={Formalizing and benchmarking prompt injection attacks and defenses},
  author={Liu, Yupei and Jia, Yuqi and Geng, Runpeng and Jia, Jinyuan and Gong, Neil Zhenqiang},
  booktitle={33rd USENIX Security Symposium (USENIX Security 24)},
  pages={1831--1847},
  year={2024}
}

@article{promptinject,
  title={Prompt injection attack against llm-integrated applications},
  author={Liu, Yi and Deng, Gelei and Li, Yuekang and Wang, Kailong and Wang, Zihao and Wang, Xiaofeng and Zhang, Tianwei and Liu, Yepang and Wang, Haoyu and Zheng, Yan and others},
  journal={arXiv preprint arXiv:2306.05499},
  year={2023}
}

@article{philosopher,
  title={The philosopher's stone: Trojaning plugins of large language models},
  author={Dong, Tian and Xue, Minhui and Chen, Guoxing and Holland, Rayne and Meng, Yan and Li, Shaofeng and Liu, Zhen and Zhu, Haojin},
  journal={arXiv preprint arXiv:2312.00374},
  year={2023}
}

@inproceedings{trojansql,
  title={Trojansql: Sql injection against natural language interface to database},
  author={Zhang, Jinchuan and Zhou, Yan and Hui, Binyuan and Liu, Yaxin and Li, Ziming and Hu, Songlin},
  booktitle={Proceedings of the 2023 Conference on Empirical Methods in Natural Language Processing},
  pages={4344--4359},
  year={2023}
}

@inproceedings{promptinjectionssqlinjection,
  title={Prompt-to-SQL injections in LLM-integrated web applications: Risks and defenses},
  author={Pedro, Rodrigo and Coimbra, Miguel E and Castro, Daniel and Carreira, Paulo and Santos, Nuno},
  booktitle={Proceedings of the IEEE/ACM 47th International Conference on Software Engineering},
  pages={1768--1780},
  year={2025}
}

@article{MASTERKEY,
  title={Masterkey: Automated jailbreak across multiple large language model chatbots},
  author={Deng, Gelei and Liu, Yi and Li, Yuekang and Wang, Kailong and Zhang, Ying and Li, Zefeng and Wang, Haoyu and Zhang, Tianwei and Liu, Yang},
  journal={arXiv preprint arXiv:2307.08715},
  year={2023}
}

@article{skillscan,
  title={Agent Skills in the Wild: An Empirical Study of Security Vulnerabilities at Scale},
  author={Liu, Yi and Wang, Weizhe and Feng, Ruitao and Zhang, Yao and Xu, Guangquan and Deng, Gelei and Li, Yuekang and Zhang, Leo},
  journal={arXiv preprint arXiv:2601.10338},
  year={2026}
}

@article{skillprobe,
  title={SkillProbe: Security Auditing for Emerging Agent Skill Marketplaces via Multi-Agent Collaboration},
  author={Guo, Zihan and Chen, Zhiyu and Nie, Xiaohang and Lin, Jianghao and Zhou, Yuanjian and Zhang, Weinan},
  journal={arXiv preprint arXiv:2603.21019},
  year={2026}
}

@article{malskill,
  title={{MalSkills}: Detecting Malicious Skills in the Agentic Supply Chain via Neuro-symbolic Reasoning},
  author={Wang, Shenao and He, Junjie and Zhao, Yanjie and Wang, Yayi and Yu, Kan and Wang, Haoyu},
  journal={arXiv preprint arXiv:2603.27204},
  year={2026}
}

@misc{mcp_filesystem,
  author       = {{Model Context Protocol}},
  title        = {{Filesystem MCP Server}},
  year         = {2025},
  howpublished = {\url{https://github.com/modelcontextprotocol/servers/blob/main/src/filesystem/README.md}},
  note         = {Accessed: 2026-04-29}
}

@misc{nist-least-privilege,
  author       = {{National Institute of Standards and Technology}},
  title        = {{Least Privilege}},
  year         = {2025},
  howpublished = {\url{https://csrc.nist.gov/glossary/term/least_privilege}},
  note         = {Accessed: 2026-04-29}
}

@misc{mcp_git,
  author       = {{Model Context Protocol}},
  title        = {{mcp-server-git}},
  year         = {2025},
  howpublished = {\url{https://pypi.org/project/mcp-server-git/}},
  note         = {Accessed: 2026-04-29}
}

@misc{mcp_terminal,
  author       = {{weidwonder}},
  title        = {{Terminal MCP Server}},
  year         = {2025},
  howpublished = {\url{https://github.com/weidwonder/terminal-mcp-server}},
  note         = {Accessed: 2026-04-29}
}

@misc{mcp_pandoc,
  author       = {{vivekVells}},
  title        = {{mcp-pandoc: MCP Server for Document Format Conversion Using Pandoc}},
  year         = {2025},
  howpublished = {\url{https://github.com/vivekVells/mcp-pandoc}},
  note         = {Accessed: 2026-04-29}
}

@misc{mcp_placeholder_image,
  author       = {{sam2332}},
  title        = {{MCP Server Placeholder Image Generator}},
  year         = {2025},
  howpublished = {\url{https://github.com/sam2332/Mcp-Server-Placeholder-Image-Generator}},
  note         = {Accessed: 2026-04-29}
}

@misc{mcp_mockloop,
  author       = {{MockLoop}},
  title        = {{MockLoop MCP Documentation}},
  year         = {2025},
  howpublished = {\url{https://docs.mockloop.com/}},
  note         = {Accessed: 2026-04-29}
}

@misc{caterpillar,
      title={Caterpillar: Caterpillar is a security scanning library for AI agent skill files (e.g., Claude Code skills) for dangerous or malicious behavior
}, 
      author={Alice-Dot-Io},
      year={2026},
      url={https://github.com/alice-dot-io/caterpillar}, 
}

@misc{nova-pro,
      title={Nova-Proximity: Nova-Proximity is a MCP and Agent Skills security scanner powered with NOVA}, 
      author={Nova-Hunting},
      year={2026},
      url={https://github.com/Nova-Hunting/nova-proximity}, 
}

@article{masb,
  title={Malicious agent skills in the wild: A large-scale security empirical study},
  author={Liu, Yi and Chen, Zhihao and Zhang, Yanjun and Deng, Gelei and Li, Yuekang and Ning, Jianting and Zhang, Ying and Zhang, Leo Yu},
  journal={arXiv preprint arXiv:2602.06547},
  year={2026}
}

@inproceedings{Agentfuzz,
  title={Make agent defeat agent: Automatic detection of $\{$Taint-Style$\}$ vulnerabilities in $\{$LLM-based$\}$ agents},
  author={Liu, Fengyu and Zhang, Yuan and Luo, Jiaqi and Dai, Jiarun and Chen, Tian and Yuan, Letian and Yu, Zhengmin and Shi, Youkun and Li, Ke and Zhou, Chengyuan and others},
  booktitle={34th USENIX Security Symposium (USENIX Security 25)},
  pages={3767--3786},
  year={2025}
}

@article{AgentRaft,
  title={AgentRaft: Automated Detection of Data Over-Exposure in LLM Agents},
  author={Lin, Yixi and Wu, Jiangrong and Nan, Yuhong and Wang, Xueqiang and Zhang, Xinyuan and Zheng, Zibin},
  journal={arXiv preprint arXiv:2603.07557},
  year={2026}
}

@article{ChainFuzzer,
  title={ChainFuzzer: Greybox Fuzzing for Workflow-Level Multi-Tool Vulnerabilities in LLM Agents},
  author={Wu, Jiangrong and Yao, Zitong and Nan, Yuhong and Zheng, Zibin},
  journal={arXiv preprint arXiv:2603.12614},
  year={2026}
}

@inproceedings{llm-as-a-judge-1,
  title     = {G-Eval: NLG Evaluation using GPT-4 with Better Human Alignment},
  author    = {Liu, Yang and Iter, Dan and Xu, Yichong and Wang, Shuohang and Xu, Ruochen and Zhu, Chenguang},
  booktitle = {Proceedings of the 2023 Conference on Empirical Methods in Natural Language Processing},
  year      = {2023}
}

@inproceedings{llm-as-a-judge-2,
  title     = {Judging LLM-as-a-Judge with MT-Bench and Chatbot Arena},
  author    = {Zheng, Lianmin and Chiang, Wei-Lin and Sheng, Ying and Zhuang, Siyuan and Wu, Zhanghao and Zhuang, Yonghao and Lin, Zi and Li, Zhuohan and Li, Dacheng and Xing, Eric P. and Zhang, Hao and Gonzalez, Joseph E. and Stoica, Ion},
  booktitle = {Advances in Neural Information Processing Systems},
  year      = {2023}
}

@inproceedings{llm-as-a-judge-3,
  title     = {LLM-Eval: Unified Multi-Dimensional Automatic Evaluation for Open-Domain Conversations with Large Language Models},
  author    = {Lin, Yen-Ting and Chen, Yun-Nung},
  booktitle = {Proceedings of the 5th Workshop on NLP for Conversational AI},
  year      = {2023}
}

@inproceedings{script-refactory-1,
  title={Em-assist: Safe automated extractmethod refactoring with llms},
  author={Pomian, Dorin and Bellur, Abhiram and Dilhara, Malinda and Kurbatova, Zarina and Bogomolov, Egor and Sokolov, Andrey and Bryksin, Timofey and Dig, Danny},
  booktitle={Companion Proceedings of the 32nd ACM International Conference on the Foundations of Software Engineering},
  pages={582--586},
  year={2024}
}

@article{script-refactory-2,
  title={Mantra: Enhancing automated method-level refactoring with contextual RAG and multi-agent LLM collaboration},
  author={Xu, Yisen and Lin, Feng and Yang, Jinqiu and Tsantalis, Nikolaos and others},
  journal={arXiv preprint arXiv:2503.14340},
  year={2025}
}

@inproceedings{script-refactory-3,
  title={Automated Extract Method Refactoring with Open-Source LLMs: A Comparative Study},
  author={Chand, Sivajeet and Kilic, Melih and W{\"u}rsching, Roland and Pandey, Sushant Kumar and Pretschner, Alexander},
  booktitle={2025 2nd IEEE/ACM International Conference on AI-powered Software (AIware)},
  pages={113--122},
  year={2025},
  organization={IEEE}
}

@misc{clawhub,
  title        = {{ClawHub}},
  author       = {{ClawHub}},
  year         = {2026},
  howpublished = {\url{https://clawhub.ai/}},
  note         = {Accessed: 2026-04-19}
}

@misc{skill-marketplace,
  title        = {{Agent Skills Marketplace}},
  author       = {{SkillsMP}},
  year         = {2026},
  howpublished = {\url{https://skillsmp.com/}},
  note         = {Accessed: 2026-04-19}
}

@misc{gpt-5-1,
  title        = {{GPT-5.1}},
  author       = {{OpenAI}},
  year         = {2025},
  howpublished = {\url{https://openai.com/index/gpt-5-1/}},
  note         = {Accessed: 2026-04-19}
}

@misc{codex,
  title        = {{Codex: AI Coding Partner from OpenAI}},
  author       = {{OpenAI}},
  year         = {2026},
  howpublished = {\url{https://openai.com/codex/}},
  note         = {Accessed: 2026-04-19}
}

@misc{claude-code,
  title        = {{Claude Code}},
  author       = {{Anthropic}},
  year         = {2026},
  howpublished = {\url{https://www.anthropic.com/product/claude-code}},
  note         = {Accessed: 2026-04-19}
}

@misc{trae,
  title        = {{TRAE: Collaborate with Intelligence}},
  author       = {{TRAE}},
  year         = {2026},
  howpublished = {\url{https://www.trae.ai/}},
  note         = {Accessed: 2026-04-19}
}

@misc{cursor,
  title        = {{Cursor: The Best Way to Code with AI}},
  author       = {{Anysphere}},
  year         = {2026},
  howpublished = {\url{https://cursor.com/}},
  note         = {Accessed: 2026-04-19}
}

@misc{gdpr,
  title        = {{General Data Protection Regulation (GDPR), Article 5: Principles relating to processing of personal data}},
  author       = {{European Union}},
  year         = {2016},
  howpublished = {\url{https://gdpr-info.eu/art-5-gdpr/}},
  note         = {Accessed: 2026-04-21}
}

@misc{data-minimization-1,
  title        = {{Principle (c): Data minimisation}},
  author       = {{Information Commissioner's Office}},
  year         = {2026},
  howpublished = {\url{https://ico.org.uk/for-organisations/uk-gdpr-guidance-and-resources/data-protection-principles/a-guide-to-the-data-protection-principles/data-minimisation/}},
  note         = {Accessed: 2026-04-21}
}

@misc{data-minimization-2,
  title        = {{Data Minimisation}},
  author       = {{European Data Protection Supervisor}},
  year         = {2026},
  howpublished = {\url{https://www.edps.europa.eu/data-protection/data-protection/glossary/d_en}},
  note         = {Accessed: 2026-04-21}
}

@inproceedings{privacy-policy-analysis,
  title     = {Polisis: Automated Analysis and Presentation of Privacy Policies Using Deep Learning},
  author    = {Harkous, Hamza and Fawaz, Kassem and Lebret, R{\'e}mi and Schaub, Florian and Shin, Kang G. and Aberer, Karl},
  booktitle = {Proceedings of the 27th USENIX Security Symposium},
  year      = {2018}
}

@inproceedings{information-flow,
  title     = {TaintDroid: An Information-Flow Tracking System for Realtime Privacy Monitoring on Smartphones},
  author    = {Enck, William and Gilbert, Peter and Chun, Byung-Gon and Cox, Landon P. and Jung, Jaeyeon and McDaniel, Patrick and Sheth, Anmol N.},
  booktitle = {Proceedings of the 9th USENIX Symposium on Operating Systems Design and Implementation},
  year      = {2010}
}

@inproceedings{privacy-leakage,
  title     = {PiOS: Detecting Privacy Leaks in iOS Applications},
  author    = {Egele, Manuel and Kruegel, Christopher and Kirda, Engin and Vigna, Giovanni},
  booktitle = {Proceedings of the Network and Distributed System Security Symposium},
  year      = {2011}
}

@article{agentdojo,
  title   = {AgentDojo: A Dynamic Environment to Evaluate Prompt Injection Attacks and Defenses for LLM Agents},
  author  = {Debenedetti, Edoardo and Zhang, Jie and Balunovi{\'c}, Mislav and Beurer-Kellner, Luca and Fischer, Marc and Tram{\`e}r, Florian},
  journal = {arXiv preprint arXiv:2406.13352},
  year    = {2024}
}

@article{ASB,
  title   = {Agent Security Bench (ASB): Formalizing and Benchmarking Attacks and Defenses in LLM-based Agents},
  author  = {Zhang, Hanrong and Huang, Jingyuan and Mei, Kai and Yao, Yifei and Wang, Zhenting and Zhan, Chenlu and Wang, Hongwei and Zhang, Yongfeng},
  journal = {arXiv preprint arXiv:2410.02644},
  year    = {2024}
}

@article{skillattack,
  title   = {SkillAttack: Automated Red Teaming of Agent Skills through Attack Path Refinement},
  author  = {Duan, Zenghao and Tian, Yuxin and Yin, Zhiyi and Pang, Liang and Deng, Jingcheng and Wei, Zihao and Xu, Shicheng and Ge, Yuyao and Cheng, Xueqi},
  journal = {arXiv preprint arXiv:2604.04989},
  year    = {2026}
}

@misc{skill-sec-scan,
  author       = {{WellAlly Technology}},
  title        = {Skill-Security-Scanner},
  year         = {2026},
  howpublished = {\url{https://github.com/huifer/skill-security-scan}},
  note         = {GitHub repository. Accessed: 2026-04-21}
}

@misc{skill-scanner,
  title        = {{Skill Scanner}},
  author       = {{Cisco AI Defense}},
  year         = {2026},
  howpublished = {\url{https://github.com/cisco-ai-defense/skill-scanner}},
  note         = {Accessed: 2026-04-21}
}

@misc{agent-skill-marketplace-security,
  title        = {Snyk Finds Prompt Injection in 36\%, 1467 Malicious Payloads in a ToxicSkills Study of Agent Skills Supply Chain Compromise},
  author       = {Luca Beurer-Kellner and Aleksei Kudrinskii and Marco Milanta and Kristian Bonde Nielsen and Hemang Sarkar and Liran Tal},
  year         = {2026},
  howpublished = {\url{https://snyk.io/blog/toxicskills-malicious-ai-agent-skills-clawhub/}},
  note         = {Accessed: 2026-04-21}
}

@article{kim2025prompt,
  title={Prompt flow integrity to prevent privilege escalation in llm agents},
  author={Kim, Juhee and Choi, Woohyuk and Lee, Byoungyoung},
  journal={arXiv preprint arXiv:2503.15547},
  year={2025}
}

@article{shi2025progent,
  title={Progent: Programmable privilege control for llm agents},
  author={Shi, Tianneng and He, Jingxuan and Wang, Zhun and Li, Hongwei and Wu, Linyu and Guo, Wenbo and Song, Dawn},
  journal={arXiv preprint arXiv:2504.11703},
  year={2025}
}

@article{jia2026skillject,
  title={Skillject: Automating stealthy skill-based prompt injection for coding agents with trace-driven closed-loop refinement},
  author={Jia, Xiaojun and Liao, Jie and Qin, Simeng and Gu, Jindong and Ren, Wenqi and Cao, Xiaochun and Liu, Yang and Torr, Philip},
  journal={arXiv preprint arXiv:2602.14211},
  year={2026}
}

@article{schmotz2026skill,
  title={Skill-inject: Measuring agent vulnerability to skill file attacks},
  author={Schmotz, David and Beurer-Kellner, Luca and Abdelnabi, Sahar and Andriushchenko, Maksym},
  journal={arXiv preprint arXiv:2602.20156},
  year={2026}
}

@article{saltzer1975protection,
  title     = {The Protection of Information in Computer Systems},
  author    = {Saltzer, Jerome H. and Schroeder, Michael D.},
  journal   = {Proceedings of the IEEE},
  volume    = {63},
  number    = {9},
  pages     = {1278--1308},
  year      = {1975},
  publisher = {IEEE},
  doi       = {10.1109/PROC.1975.9939}
}

@techreport{nist80053,
  title       = {Security and Privacy Controls for Information Systems and Organizations},
  author      = {{National Institute of Standards and Technology}},
  institution = {National Institute of Standards and Technology},
  type        = {Special Publication},
  number      = {800-53 Rev. 5},
  year        = {2020},
  note        = {Control AC-6: Least Privilege}
}

@inproceedings{felt2011android,
  title     = {Android Permissions Demystified},
  author    = {Felt, Adrienne Porter and Chin, Erika and Hanna, Steve and Song, Dawn and Wagner, David},
  booktitle = {Proceedings of the 18th ACM Conference on Computer and Communications Security (CCS)},
  pages     = {627--638},
  year      = {2011},
  publisher = {ACM},
  doi       = {10.1145/2046707.2046779}
}

@inproceedings{roesner2012user,
  title     = {User-Driven Access Control: Rethinking Permission Granting in Modern Operating Systems},
  author    = {Roesner, Franziska and Kohno, Tadayoshi and Moshchuk, Alexander and Parno, Bryan and Wang, Helen J. and Cowan, Crispin},
  booktitle = {Proceedings of the 2012 IEEE Symposium on Security and Privacy (S\&P)},
  pages     = {224--238},
  year      = {2012},
  publisher = {IEEE},
  doi       = {10.1109/SP.2012.24}
}

@inproceedings{shuster2021retrieval,
  title     = {Retrieval Augmentation Reduces Hallucination in Conversation},
  author    = {Shuster, Kurt and Poff, Spencer and Chen, Moya and Kiela, Douwe and Weston, Jason},
  booktitle = {Findings of the Association for Computational Linguistics: EMNLP 2021},
  pages     = {3784--3803},
  year      = {2021},
  address   = {Punta Cana, Dominican Republic},
  publisher = {Association for Computational Linguistics},
  doi       = {10.18653/v1/2021.findings-emnlp.320},
  url       = {https://aclanthology.org/2021.findings-emnlp.320/}
}

@inproceedings{lewis2020retrieval,
  title     = {Retrieval-Augmented Generation for Knowledge-Intensive {NLP} Tasks},
  author    = {Lewis, Patrick and Perez, Ethan and Piktus, Aleksandra and Petroni, Fabio and Karpukhin, Vladimir and Goyal, Naman and K{\"u}ttler, Heinrich and Lewis, Mike and Yih, Wen-tau and Rockt{\"a}schel, Tim and Riedel, Sebastian and Kiela, Douwe},
  booktitle = {Advances in Neural Information Processing Systems},
  volume    = {33},
  pages     = {9459--9474},
  year      = {2020},
  publisher = {Curran Associates, Inc.},
  url       = {https://proceedings.neurips.cc/paper/2020/hash/6b493230-Abstract.html}
}

@inproceedings{bechard2024reducing,
  title     = {Reducing Hallucination in Structured Outputs via Retrieval-Augmented Generation},
  author    = {B{\'e}chard, Patrice and Marquez Ayala, Orlando},
  booktitle = {Proceedings of the 2024 Conference of the North American Chapter of the Association for Computational Linguistics: Human Language Technologies (Volume 6: Industry Track)},
  pages     = {228--238},
  year      = {2024},
  address   = {Mexico City, Mexico},
  publisher = {Association for Computational Linguistics},
  doi       = {10.18653/v1/2024.naacl-industry.19},
  url       = {https://aclanthology.org/2024.naacl-industry.19/}
}

@inproceedings{yao2023react,
  title     = {{ReAct}: Synergizing Reasoning and Acting in Language Models},
  author    = {Yao, Shunyu and Zhao, Jeffrey and Yu, Dian and Du, Nan and Shafran, Izhak and Narasimhan, Karthik and Cao, Yuan},
  booktitle = {The Eleventh International Conference on Learning Representations},
  year      = {2023},
  url       = {https://openreview.net/forum?id=WE_vluYUL-X}
}

\section*{Ethical Considerations}

Our study analyzes publicly available Agent Skill bundles from public repositories and marketplaces, and does not collect private user data, execute Skills in real user environments, or access private services. All dynamic executions, including original runs, replay runs, and privilege-constraining validation, are conducted in a sandbox that isolates filesystem access, network communication, command execution, and other external side effects. For potentially sensitive artifacts such as credentials, local environment files, command histories, or external endpoints, we block the operation or retain only redacted evidence, and we do not intentionally trigger real data exfiltration, destructive operations, or unauthorized interactions. To reduce dual-use risks, we release only sanitized examples, aggregate statistics, and non-sensitive reproduction materials. Our goal is to support Skill marketplace governance, pre-installation checking, and user-side risk reduction through fine-grained data-minimization and least-privilege enforcement.



\appendix

\section{Quality of Unified Execution Graph Construction}
\label{app:graph_quality}

\begin{table}[tbhp]
\centering
\caption{The quality of unified execution graph components. TN is omitted because graph extraction does not have a naturally bounded negative space.}
\label{tab:graph_quality}
\resizebox{\linewidth}{!}{%
\begin{tabular}{l|c|c|c|c}
\hline
\textbf{Component} & \textbf{TP/FP/FN} & \textbf{Precision} & \textbf{Recall} & \textbf{F1} \\
\hline
Instruction nodes   & 1441/32/25 & 97.84 & 98.31 & 98.07 \\
Instruction edges   & 1480/47/40 & 96.92 & 97.38 & 97.15 \\
Code nodes          & 617/35/30  & 94.63 & 95.41 & 95.02 \\
Code edges          & 765/65/57  & 92.18 & 93.06 & 92.62 \\
Cross-layer links   & 194/10/8   & 95.11 & 96.04 & 95.57 \\
\hline
\end{tabular}%
}
\end{table}

Since \system{} is fundamentally built on a graph-based analysis pipeline, the quality of the unified execution graph directly affects the reliability of subsequent candidate extraction, task instantiation, replay-based validation, and privilege constraining. We therefore evaluate the quality of the constructed unified execution graph by manually annotating the major graph components, including instruction nodes, instruction edges, code nodes, code edges, and cross-layer links.

The results are shown in \Cref{tab:graph_quality}. Overall, the graph construction quality is high across all evaluated components. At the instruction level, \system{} achieves 98.07\% F1 for instruction nodes and 97.15\% F1 for instruction edges, indicating that it can accurately recover the high-level procedural structure expressed in natural-language skill instructions. At the code level, \system{} achieves 95.02\% F1 for code nodes and 92.62\% F1 for code edges, showing that the extracted code graph can reliably capture concrete operation-level behaviors and their dependencies. For cross-layer links, which connect instruction-level invocations with code-level execution units, \system{} achieves 95.57\% F1, suggesting that the composed unified graph can effectively preserve the interactions between high-level instructions and low-level scripts.

These results show that the unified execution graph used by \system{} is sufficiently accurate to serve as the structural foundation of our graph-based least-privilege analysis. Although code edges are slightly harder to recover than other components due to more complex control/data dependencies in scripts, the overall precision and recall remain consistently high, supporting the effectiveness of downstream graph-guided analysis.

\begin{table*}[t]
\centering
\caption{MCP-backed resource fixtures supported during graph-guided task instantiation.}
\label{tab:resource-fixtures}
\small
\setlength{\tabcolsep}{3.5pt}
\renewcommand{\arraystretch}{1.15}
\begin{tabular}{p{0.16\textwidth}|p{0.22\textwidth}|p{0.43\textwidth}|p{0.13\textwidth}}
\Xhline{1.1pt}
\textbf{Task Type} 
& \textbf{Typical Requirement} 
& \textbf{MCP-backed Fixture Strategy} 
& \textbf{MCP Server} \\
\hline
Prompt-only task
& Status checking, local summary, or simple inspection.
& No external fixture is required.
& - \\
\hline
File-based task
& CSV, JSON, TXT, or LOG input.
& \system{} creates a minimal synthetic file with expected fields and sample values using file-write and directory-creation operations.
& Filesystem~\cite{mcp_filesystem} \\
\hline
Repo-based task
& Repository analysis, commit inspection, or diff summarization.
& \system{} creates a toy repository fixture with a few files and commits. File contents are created through filesystem operations, while repository operations are performed through Git tools. If repository initialization is required, \system{} uses a sandboxed command-execution MCP\@.
& Filesystem~\cite{mcp_filesystem}; Git~\cite{mcp_git}; Terminal~\cite{mcp_terminal} \\
\hline
Document-based task
& Markdown, TXT, DOCX, or PDF-style document input.
& \system{} creates a minimal Markdown/TXT document directly, or converts a simple Markdown/HTML/TXT source into DOCX/PDF when the Skill expects richer document formats.
& Filesystem~\cite{mcp_filesystem}; Pandoc~\cite{mcp_pandoc} \\
\hline
Image-based task
& Image inspection, conversion, or metadata-related behavior.
& \system{} creates a small placeholder PNG image with configurable size/color and stores it at the expected path.
& Placeholder Image~\cite{mcp_placeholder_image} \\
\hline
Config/API-based task
& Configuration loading, endpoint access, or API-style interaction.
& \system{} creates a dummy configuration file through filesystem operations. If an executable mock endpoint is required, \system{} instantiates a mock API server from a lightweight OpenAPI specification.
& Filesystem~\cite{mcp_filesystem}; MockLoop~\cite{mcp_mockloop}  \\
\Xhline{1.1pt}
\end{tabular}
\end{table*}

\section{Resource Fixtures for Task Instantiation}
\label{app:resource-fixtures}
As shown in \Cref{tab:resource-fixtures}, for resource-dependent Skills, \system{} augments generated user prompts with lightweight synthetic fixtures. These fixtures are not intended to model all real-world inputs but to provide minimal executable inputs for triggering the intended graph chain during validation.

\section{Dataset Construction for Effectiveness Evaluation}
\label{app:dataset-construction}

To improve the representativeness and reliability of the manually labeled evaluation set, we construct the 200-Skill dataset through a process combining stratified sampling with multi-stage annotation. The key idea is to build ground truth from the original Skill artifacts and manually verified task contexts, rather than labeling solely at the Skill level.

\para{Sampling Strategy}
We sample Skills from the valid ecosystem-scale corpus used for large-scale measurement, i.e., Skills that remain available after deduplication and availability filtering. To avoid concentrating the benchmark on a narrow subset of the ecosystem, we stratify the sampling process along four dimensions: marketplace source~\cite{clawhub, skill-marketplace}, Skill category, artifact composition, and observable operation type. Artifact composition covers instruction-only Skills and Skills that contain both natural-language instructions and executable scripts. Observable operation types include file access, network communication, external transmission, command execution, and persistent state modification. These criteria help the evaluation set cover diverse Skill structures and behavior surfaces.

\para{Ground-truth Graph and Task Construction}
For each sampled Skill, we inspect the original Skill artifacts, including metadata, instruction files, scripts, and auxiliary resources, and construct a ground-truth unified execution graph. The graph contains instruction-level action nodes, code-level action nodes, predicate nodes, and execution dependency edges, including control, data, call, and return edges. Each action node is recorded with its normalized operation, target object, source layer, and provenance in the original artifact. The graph is manually checked and revised until it faithfully represents the Skill behaviors exposed by the inspected artifacts. Based on this verified graph, we extract representative execution chains corresponding to the Skill's supported task variants. For each chain, we generate and refine a user prompt and verify that the prompt is semantically consistent with the Skill's declared functionality and that it structurally triggers the intended chain. For resource-dependent tasks, we also verify that the prompt and its required fixture together exercise the intended chain. The manually verified graph and prompts are used only for constructing benchmark ground truth, not as inputs to \system{} during evaluation.


\hypertarget{rev:m1-annotation-criteria}{}\para{Action-in-task Labeling and Aggregation}
Ground truth is assigned at the action-in-task level. For each Skill $\sigma$, privilege-relevant action $a$, and verified task $p$, annotators determine whether the operation, data source, sink or destination, and side effect of $a$ are authorized by $p$. Independently, they assess the original/modified execution pair to determine whether neutralizing $a$ preserves the prompt-required core flow and whether the modified execution still satisfies the user's goal, corresponding to $\mathsf{CorePres}_{p}(\tau^{original},\tau^{replay})$ and $\mathsf{GoalSat}_{p}(\tau^{original},\tau^{replay})$. The pair is labeled over-privileged when $\neg\mathsf{Auth}_{p}(a)\vee\mathsf{Unnec}_{\sigma}(a,p)$ holds. If the final verdict cannot be resolved from the inspected evidence, the pair is conservatively treated as not confirmed. We then aggregate action-in-task labels into action-level and Skill-level labels: an action is labeled over-privileged if it has at least one positive verified action-in-task verdict, and a Skill is labeled over-privileged if it contains at least one over-privileged action-in-task instance. A Skill is labeled benign when no over-privileged action-in-task instance is found under the annotated task set. This aggregation provides ground truth for the three evaluation granularities used in \Cref{tab:rq2_end2end}: action-in-task, action-level, and Skill-level.


\para{Annotation Quality Control and Final Set}
Each Skill is independently inspected by two annotators. Disagreements on graph structure, task-chain coverage, prompt validity, action boundaries, task authorization, core-flow preservation, goal satisfaction, or the final action verdict are resolved through discussion.
Cases without annotator consensus are reviewed by a third annotator.
If, after third-annotator review, the concrete prompt, action, original/replay traces, and outputs still provide insufficient evidence to establish the task-conditioned over-privilege criterion in \Cref{subsec:problem statement}, the action-in-task pair is conservatively treated as not confirmed rather than labeled over-privileged.
The final benchmark contains 200 Skills, including 100 Skills with over-privileged actions and 100 benign Skills without over-privileged actions under the annotated task set. At the graph level, the benchmark contains 2,113 annotated action nodes, including 289 over-privileged action nodes and 1,824 non-over-privileged action nodes. At the task level, annotators identify 261 ground-truth task chains. These triggered tasks produce 737 annotated action-in-task instances, including 507 over-privileged instances and 230 non-over-privileged instances. This evaluation set is used for effectiveness and privilege-constraining evaluation.

\section{Token Cost Analysis}
\label{app:token_cost}

To evaluate the computational cost of \system{}, we measure the LLM token usage on the 200-skill benchmark used in \Cref{subsec:rq1}. During the evaluation, we instrument all LLM API calls issued by \system{} and record their input tokens, output tokens, and call counts. Since \system{} is designed for offline ecosystem-side analysis, we focus on token usage and LLM-call counts. The detailed results are shown in \Cref{tab:token_cost_details}. All token counts are reported in millions (M).

\begin{table*}[t]
\centering
\caption{Token cost of \system{} on the 200-skill dataset.}
\label{tab:token_cost_details}
\resizebox{\textwidth}{!}{%
\renewcommand{\arraystretch}{1.1}
\fontsize{8pt}{10pt}\selectfont
\begin{tabular}{l|c|c|c|c|c|c}
\Xhline{0.7pt}
\textbf{Module / Stage}
& \makecell[c]{\textbf{Processed}\\\textbf{Unit}}
& \makecell[c]{\textbf{\#}\\\textbf{Units}}
& \makecell[c]{\textbf{\# LLM}\\\textbf{Calls}}
& \makecell[c]{\textbf{Input}\\\textbf{Tokens (M)}}
& \makecell[c]{\textbf{Output}\\\textbf{Tokens (M)}}
& \makecell[c]{\textbf{Avg. Tokens}\\\textbf{/ Unit (M)}} \\
\hline
\rule{0pt}{2.2ex}
Unified Execution Graph Construction
& Skill
& 200
& 200
& 0.358
& 0.111
& 0.00235 \\

\rule{0pt}{2.2ex}
Action Consistency Screening
& Action node
& 2,113
& 2,113
& 2.217
& 0.170
& 0.00113 \\

\rule{0pt}{2.2ex}
User Task/Prompt Instantiation
& Action chain
& 261
& 261
& 0.208
& 0.021
& 0.00088 \\

\rule{0pt}{2.2ex}
Over-Privilege Validation
& Action-in-task case
& 737
& 2,211
& 4.205
& 0.184
& 0.00596 \\

\rule{0pt}{2.2ex}
Control-flow Privilege Constraining
& Repaired action
& 279
& 666
& 1.304
& 0.558
& 0.00667 \\ \hline

\textbf{Overall}
& Skill
& \textbf{200}
& \textbf{5,451}
& \textbf{8.292}
& \textbf{1.045}
& \textbf{0.04669} \\
\Xhline{0.7pt}
\end{tabular}%
}
\end{table*}

\para{Stage-wise Token Cost}
We attribute each recorded LLM call to the corresponding module of \system{}. In the over-privilege candidate extraction stage, token usage comes from unified execution graph construction and action consistency screening. The former normalizes instruction-level procedures and recovers semantic dependencies, while the latter performs node-level consistency judgments over extracted action nodes. In the action over-privilege validation stage, token usage comes from user task/prompt instantiation and over-privilege validation. The prompt instantiation module generates user tasks from graph-represented action chains, while over-privilege validation compares original and replay executions under concrete task contexts. Finally, in the privilege constraining stage, token usage comes from repair planning, guard synthesis, and projection of the constrained behavior back to the original skill artifacts.

\para{Overall Cost}
Overall, \system{} issues 5,451 LLM calls on the 200-skill benchmark, consuming 8.292M input tokens and 1.045M output tokens. This corresponds to 0.04669M total tokens per skill on average. The largest token cost comes from over-privilege validation, which consumes 4.205M input tokens and 0.184M output tokens. This is expected because the validation needs to reason over concrete user tasks, candidate actions, original/replay traces, and final outputs. Control-flow privilege constraining has the highest average cost per repaired action, because it requires generating concrete repair plans and projecting them back to instruction or code artifacts. These results show that the main cost of \system{} comes from semantic validation and repair, while graph construction and task instantiation remain relatively lightweight.

\begin{table*}[t]
\centering
\caption{Post-hoc manual validation for calibrating large-scale measurement at the skill, unified-graph, and user-task levels.}
\label{tab:posthoc_human_audit}
\resizebox{\textwidth}{!}{%
\renewcommand{\arraystretch}{1}
\fontsize{8pt}{10pt}\selectfont
\begin{tabular}{l|c|c|c|c|c|c|c}
\Xhline{0.7pt}
\textbf{Audit Target}
& \makecell[c]{\textbf{Population}\\\textbf{Size}}
& \makecell[c]{\textbf{Sample}\\\textbf{Size}}
& \makecell[c]{\textbf{Instruction-only}\\\textbf{Samples}}
& \makecell[c]{\textbf{Code-involving}\\\textbf{Samples}}
& \makecell[c]{\textbf{Human-confirmed}\\\textbf{Valid Claims}}
& \makecell[c]{\textbf{Weighted}\\\textbf{Validity Rate}}
& \makecell[c]{\textbf{95\%}\\\textbf{CI}} \\
\hline
\rule{0pt}{2.2ex}
\makecell[l]{Skill-level validated\\over-privilege}
& \makecell[c]{6,590\\(5,092 / 1,498)}
& 300
& 200
& 100
& \makecell[c]{189 / 200\\93 / 100}
& 94.16\%
& [91.47, 96.85] \\ \hline

\makecell[l]{Over-privilege\\candidate action nodes}
& \makecell[c]{11,588\\(8,417 / 3,171)}
& 200
& 150
& 50
& \makecell[c]{140 / 150\\46 / 50}
& 92.97\%
& [89.41, 96.52] \\ \hline

\makecell[l]{User-task-level validated\\over-privilege cases}
& \makecell[c]{7,033\\(5,391 / 1,642)}
& 200
& 150
& 50
& \makecell[c]{141 / 150\\46 / 50}
& 93.53\%
& [90.13, 96.93] \\ \hline
\Xhline{0.7pt}
\end{tabular}%
}
\end{table*}

\section{Post-hoc Manual Validation of Large-scale Measurement}
\label{app:human_audit}

To further calibrate the reliability of the large-scale measurement results in \Cref{tab:rq1_overall}, we conduct a post-hoc human audit of the major claims reported by \system{}. The results of this audit are shown in \Cref{tab:posthoc_human_audit}. This audit is not intended to re-evaluate the end-to-end effectiveness of \system{}, which has already been measured on the independently labeled benchmark in \Cref{subsec:rq1}. Instead, its goal is to provide a manual sanity check on whether the large-scale statistics reported in the wild are supported by human-checkable evidence.


\para{Audit Granularity}
We audit the large-scale results at three granularities, corresponding to the three levels reported in \Cref{tab:rq1_overall}. First, at the \emph{skill level}, we audit validated over-privileged skills. Each sampled unit is a skill-level evidence package, and the reviewer determines whether the skill contains at least one valid task-conditioned over-privilege case. Second, at the \emph{action level}, we audit over-privilege candidate action nodes. Each sampled unit is an action node together with its provenance, upstream context, and skill profile, and the reviewer determines whether the node is a valid suspicious high-impact candidate. Third, at the \emph{user-task level}, we audit validated over-privilege cases under concrete user tasks. Each sampled unit is a task-action evidence chain, including the user task, the reported over-privileged action, original/replay traces, and original/replay outputs.

\hypertarget{rev:statistics-audit-strata}{}\para{Sampling Strategy}
We use source-aware stratified sampling to cover both instruction-level and code-level behaviors. To keep the sampling strata mutually exclusive, any unit with code-level provenance, including a unit jointly induced by instructions and code, is assigned to the \emph{code-involving} stratum; the remaining units form the \emph{instruction-only} stratum. Since code-involving cases are fewer but often more complex to inspect, we mildly oversample them to ensure sufficient coverage. Specifically, for the skill-level audit, we sample 300 validated over-privileged skills, including 200 instruction-only and 100 code-involving samples. For the graph-level audit, we sample 200 candidate action nodes, including 150 instruction-only and 50 code-involving samples. For the user-task-level audit, we sample 200 validated task-level cases, including 150 instruction-only and 50 code-involving samples. This design allows the audit to cover all major reported claim types without requiring exhaustive re-labeling of all action-task combinations.

\para{Audit Procedure}
For each sampled unit, we inspect the corresponding evidence package produced by \system{}. For skill-level samples, the package includes the skill metadata, relevant \texttt{SKILL.md} span, relevant code span if applicable, the reported over-privileged action, the generated user task, original and replay traces, original and replay outputs, and the explanation generated by \system{}. For graph-level samples, the package includes the candidate action node, its normalized operation and object, provenance span, upstream graph context, and the declared skill profile. For user-task-level samples, the package includes the concrete user task, the candidate action, the original execution, the replay execution with the candidate removed or neutralized, and the corresponding outputs. Each sampled claim is manually reviewed according to the definitions in \Cref{subsec:problem statement}. Ambiguous cases are conservatively treated as unconfirmed unless the evidence clearly supports the reported claim.

\para{Audit Criteria}
A skill-level claim is valid when the sampled Skill contains a triggered privilege-relevant action that satisfies $\neg\mathsf{Auth}_{p}(a)\vee\mathsf{Unnec}_{\sigma}(a,p)$ under a concrete task. A graph-level claim is marked as valid if the sampled action node corresponds to a high-impact or privilege-relevant behavior that is plausibly detached from, excessive for, or security-sensitive with respect to the declared skill functionality, and therefore should be forwarded for task-conditioned validation. A user-task-level claim is valid when the reported action is outside the authority conveyed by the prompt or when neutralizing it preserves the prompt-required core flow and leaves the replay result satisfying the user's goal.

\hypertarget{rev:statistics-audit-weighting}{}\para{Weighted Validity Rate}
Because code-involving cases are oversampled relative to their population size, we report population-weighted validity rates. For each audit target, let $N_I$ and $N_C$ denote the mutually exclusive instruction-only and code-involving population sizes, $n_I$ and $n_C$ denote the corresponding sample sizes, and $h_I$ and $h_C$ denote the number of manually confirmed valid claims. Thus, $N_I+N_C$ equals the deduplicated population size reported in \Cref{tab:posthoc_human_audit}. We first compute the validity rate within each source stratum:
\[
\hat{p}_I=\frac{h_I}{n_I}, \quad
\hat{p}_C=\frac{h_C}{n_C}.
\]
The overall weighted validity rate is then computed as
\[
\hat{p}
=
\frac{N_I}{N_I+N_C}\hat{p}_I
+
\frac{N_C}{N_I+N_C}\hat{p}_C.
\]
We compute the 95\% confidence interval using the standard stratified-sampling variance:
\[
\widehat{\mathrm{SE}}(\hat{p})
=
\sqrt{
\left(\frac{N_I}{N_I+N_C}\right)^2
\frac{\hat{p}_I(1-\hat{p}_I)}{n_I}
+
\left(\frac{N_C}{N_I+N_C}\right)^2
\frac{\hat{p}_C(1-\hat{p}_C)}{n_C}
}.
\]
The 95\% confidence interval is reported as
\[
\hat{p} \pm 1.96 \cdot \widehat{\mathrm{SE}}(\hat{p}).
\]
This confidence interval captures sampling uncertainty under the stratified audit design; it does not model potential reviewer subjectivity.

\section{Additional Details of Over-Privileged Skills in the Wild}
\label{app:rq2_details}

\begin{table}[thbp]
\centering
\caption{Category distribution of skills with validated over-privilege actions.}
\label{tab:rq1_category}
\resizebox{\linewidth}{!}{%
\renewcommand{\arraystretch}{2}
\fontsize{20pt}{14pt}\selectfont
\begin{tabular}{l r | l r}
\Xhline{1.5pt}
\textbf{Category} & \textbf{Count} & \textbf{Category} & \textbf{Count} \\
\hline

tools              & 2,211 / 6,590 (33.55\%) & cicd               & 340 / 6,590  (5.16\%) \\
productivity-tools & 1,608 / 6,590 (24.40\%) & testing-security   & 308 / 6,590  (4.67\%) \\
llm-ai             & 750 / 6,590  (11.38\%)   & finance-investment & 277 / 6,590 (4.20\%) \\
data-ai            & 410 / 6,590  (6.22\%)    & other              & 330 / 6,590  (5.01\%) \\
devops             & 356 / 6,590  (5.40\%)    &                    &            \\
\Xhline{1.5pt}
\end{tabular}%
}
\end{table}

\para{Category Distribution}
\Cref{tab:rq1_category} further shows that validated over-privileged behaviors are concentrated in several capability-rich categories. In particular, tools account for the largest share, with 2,211 skills (33.55\%), followed by productivity-tools with 1,608 skills (24.40\%), and llm-ai with 750 skills (11.38\%). Other categories, such as data-ai and devops, also contribute a non-trivial number of cases.

These categories typically expose broader capability surfaces, such as local resource access, external service interaction, workflow automation, and action chaining. As a result, they provide more opportunities for skills to perform unnecessary high-impact actions beyond the user’s intended task scope.

\section{Deployment on Real-World Agent Runtimes}
\label{app:agent_setup}

To evaluate whether the constrained skills produced by \system{} remain effective beyond our local sandbox, we deploy them on four representative real-world agent runtimes: Claude Code~\cite{claude-code}, Codex~\cite{codex}, Cursor~\cite{cursor}, and Trae~\cite{trae}. We use the same 95 over-privileged skills evaluated in \Cref{tab:rq3_overall_repair}. For each skill, we deploy the constrained skill bundle generated by \system{} without manually editing its repaired instructions or scripts. The same user prompt set is used across all runtimes.

\para{Evaluation Inputs}
The original baseline is fixed at 279 triggered over-privileged actions because these are the over-privileged actions observed in the 95 repair targets before constraining, as reported in \Cref{tab:rq3_overall_repair}. For each real-world agent, we execute the constrained versions of the same 95 skills and measure how many of these originally triggered over-privileged actions remain triggerable. Therefore, \Cref{tab:rq3_agent_repair} evaluates the transferability of the same repairs across different agent runtimes, rather than selecting only locally successful repair cases.

\para{Permission and Execution Settings}
We evaluate all agents under the same controlled execution policy. Filesystem access is restricted to an isolated workspace containing the test repository and skill resources. Network access and command execution are enabled only inside the controlled evaluation environment, so that over-privileged behaviors can be observed without affecting external systems. When an agent runtime provides native permission prompts or confirmation dialogs, we use a consistent non-interactive evaluation setting that pre-authorizes the sandboxed operations required by the test. This avoids measuring differences in human permission-interaction policies and instead focuses the evaluation on whether the constrained skill itself suppresses over-privileged behavior.
This platform-level test permission does not establish $\mathsf{Auth}_{p}(a)$, which is determined only from the authority conveyed by the concrete user task.

\section{Details of Task-conditioned Constrained Node Generation}
\label{app:constrained-node-generation}


This appendix provides additional details on how \system{} derives a task-conditioned constrained node from final validation verdicts. The goal is to avoid binding a constraint to a single prompt, while still preventing a validated over-privileged action from being triggered under task contexts where the final verdict marks it as over-privileged.

\para{Procedure}
For each validated action $a=\langle \ell,op,obj\rangle$, \system{} derives the constrained control node through the following steps. \Cref{tab:task-context-descriptor} summarizes the fixed-schema descriptor used in this process.

\begin{enumerate}[leftmargin=*]
    \item \textbf{Descriptor extraction.}
    \system{} collects the generated user tasks used during validation and extracts a fixed-schema task-context descriptor from each task. Each descriptor records the task intent, requested operation, target object, execution scope, any external destination, whether the user explicitly requests the side effect corresponding to $a$, and the final over-privilege verdict for $a$ under that task.

    \item \textbf{Slot normalization.}
    \system{} normalizes semantically equivalent fields in the descriptors. For example, expressions such as ``send'', ``sync'', and ``upload'' are mapped to the same external-transmission operation when they refer to the same object and destination.

    \item \textbf{Task-context grouping.}
    \system{} groups descriptors with equivalent task semantics into task-context clusters. Each cluster represents a reusable task context rather than a single prompt.

    \item \textbf{Guard synthesis.}
    Based on the clustered contexts, \system{} derives an allow condition $C_a$ that permits $a$ only when the current user task explicitly requires the operation, object, scope, and side effect represented by $a$.

    \item \textbf{Constrained node generation.}
    Finally, \system{} creates the constrained control node $v_a^c=\langle C_a,a\rangle$. During execution, $v_a^c$ routes control to $a$ only when the current task satisfies $C_a$; otherwise, the guarded action is skipped.
\end{enumerate}

\begin{table*}[t]
\centering
\caption{Fixed-schema task-context descriptor used for constrained node generation.}
\label{tab:task-context-descriptor}
\resizebox{\linewidth}{!}{%
\renewcommand{\arraystretch}{1.15}
\begin{tabular}{l|p{0.62\linewidth}}
\Xhline{1.1pt}
\textbf{Field} & \textbf{Description} \\
\hline
Task intent & The high-level goal expressed by the user task, e.g., local report generation or external report synchronization. \\ \hline
Requested operation & The operation explicitly requested by the user, e.g., generate, summarize, read, send, sync, or execute. \\ \hline
Target object & The object involved in the requested operation, e.g., report, local records, repository commits, or environment files. \\ \hline
Execution scope & Whether the requested behavior is local-only, cross-resource, or externally visible. \\ \hline
Destination & The external recipient or endpoint, if the task explicitly specifies one. \\ \hline
Explicit side-effect request & Whether the user explicitly requests the side effect corresponding to the action $a$, such as data transmission or command execution. \\ \hline
Final validation verdict & Whether $a$ is validated as over-privileged or not over-privileged for task $p$ under the final decision rule in \Cref{subsec:problem statement}; unresolved verdicts are excluded from automatic synthesis. \\ \hline
\Xhline{1.1pt}
\end{tabular}}
\end{table*}

\para{Worked Example}
Consider a validated action $a=\textsf{send}(r,t)$, where $r=\textit{report}$ and $t=\textit{Telegram}$.
During validation, \system{} may generate several user tasks involving the same Skill. \Cref{tab:task-context-example} shows how these tasks are converted into task-context descriptors and grouped according to their relation to $a$.

\begin{table*}[t]
\centering
\caption{Example task-context grouping for $a=\textsf{send}(r,t)$.}
\label{tab:task-context-example}
\resizebox{\linewidth}{!}{%
\renewcommand{\arraystretch}{1.15}
\begin{tabular}{p{0.36\linewidth}|p{0.25\linewidth}|p{0.25\linewidth}}
\Xhline{1.1pt}
\textbf{User Task} & \textbf{Extracted Context} & \textbf{Relation to $a$} \\ 
\hline
Generate a local report for my deep-work history. 
& Local report generation; object: report; scope: local-only; destination: none.
& Does not require external transmission.
Final verdict: over-privileged; rejecting context. \\
\hline
Show my deep-work heatmap locally. 
& Local visualization; object: heatmap/report; scope: local-only; destination: none.
& Does not require external transmission.
Final verdict: over-privileged; rejecting context. \\
\hline
Generate the report and send it to Alex on Telegram. 
& External report synchronization; object: report; scope: external; destination: Telegram.
& Explicitly requires sending the report.
Final verdict: not over-privileged; allowing context. \\
\hline
Sync the report to the configured Telegram recipient.
& External report synchronization; object: report; scope: external; destination: Telegram.
& Explicitly requires sending the report.
Final verdict: not over-privileged; allowing context. \\
\Xhline{1.1pt}
\end{tabular}}
\end{table*}

From these descriptors, \system{} forms two task-context clusters.
The first cluster contains local-report tasks, where the user only requests local generation or visualization and does not request any external transmission. 
The second cluster contains external-sync tasks, where the user explicitly requests sending or synchronizing the report to Telegram.
Therefore, \system{} derives an allow condition $C_a$ that permits the send action only in the external-sync context:
\[
\begin{aligned}
C_a:\quad
&\textsf{explicit\_request}(\textsf{send/sync}) \\
&\wedge\ \textsf{object}=r \\
&\wedge\ \textsf{destination}=t.
\end{aligned}
\]
The resulting constrained node is:
\[
\begin{aligned}
v_a^c
= \left\langle
C_a,\ 
\textsf{send}(r,t)
\right\rangle.
\end{aligned}
\]
During execution, if the current task only asks for local report generation, $C_a$ is not satisfied and the send action is skipped. 
If the current task explicitly asks to send or synchronize the report to Telegram, $C_a$ is satisfied and the guarded action remains reachable.

\end{document}